\documentclass{aa}  
\usepackage{graphicx}
\usepackage{txfonts}
\usepackage{natbib}
\newcommand{\bz}{\ensuremath{\langle B_z\rangle}}
\newcommand{\bs}{\ensuremath{\langle \vert B \vert \rangle}}

\newcommand{\te}{\ensuremath{T_{\mathrm{eff}}}}
\newcommand{\mwda}{{\rm WD\,2047+372}}
\newcommand{\mwdb}{{\rm WD\,2359-434}}
\newcommand{\esp}{{\rm ESPaDOnS}}
\newcommand{\ha}{{\rm H$\alpha$}}
\newcommand{\snr}{\ensuremath{S/N}}
\begin{document}

\title{Monitoring and modelling of white dwarfs with extremely weak magnetic fields
\thanks{Based in part observations collected at the European
Organisation for Astronomical Research in the Southern Hemisphere,
Chile, under observing programmes 095.D-0264 and 097.D-0264, and
obtained from the ESO/ST-ECF Science Archive Facility; in part on
observations made with the William Herschel Telescope, operated on the
island of La Palma by the Isaac Newton Group in the Spanish
Observatorio del Roque de los Muchachos of the Instituto de
Astrofisica de Canarias; and in part on observations obtained at the
Canada-France-Hawaii Telescope (CFHT) which is operated by the
National Research Council of Canada, the Institut National des
Sciences de l'Univers of the Centre National de la Recherche
Scientifique of France, and the University of Hawaii.}}

    \subtitle{WD\,2047+372 and WD\,2359$-$434}

    \titlerunning{Short-period magnetic variability of two super-weak
field white dwarfs}

   \author{J. D. Landstreet
          \inst{1,2}
          \and
          S. Bagnulo 
	  \inst{1}
	  \and
	  G. Valyavin
          \inst{3}
          \and
          A. F. Valeev
          \inst{3}
          }

   \institute{Armagh Observatory and Planetarium, College Hill, Armagh BT61 9DG,
              Northern Ireland, UK
         \and
              Dept.\ of Physics \& Astronomy, University of Western 
              Ontario, London, Ontario, Canada 
              \email{jlandstr@uwo.ca}
         \and
              Special Astrophysical Observatory, RAS, Nizhnij Arkhiz, 
              Zelenchukskij Region, 369167 Karachai-Cherkessian Republic,
              Russia
             }

   \date{Received July 01, 2017; accepted July 01, 2016}

\abstract{Magnetic fields are detected in a few percent of white dwarfs.
The number of such magnetic white dwarfs known is now some
hundreds. Fields range in strength from a few kG to several hundred
MG. Almost all the known magnetic white dwarfs have a mean field
modulus $\ge 1$\,MG. We are trying to fill a major gap in
observational knowledge at the low field limit ($\le 200$\,kG) using
circular spectro-polarimetry. 

In this paper we report the discovery and monitoring of strong,
periodic magnetic variability in two previously
discovered ``super-weak field'' magnetic white dwarfs, \mwda\ and \mwdb.
\mwda\ has a mean longitudinal field that reverses between about $-12$
and $+15$\,kG, with a period of 0.243\,d, while its mean field modulus
appears nearly constant at 60\,kG. The observations can be
intepreted in terms of a dipolar field tilted with respect to the
stellar rotation axis. 
\mwdb\ always shows a weak positive longitudinal field with values
between about 0 and $+12$\,kG, varying only weakly with stellar
rotation, while the mean field modulus varies between about 50 and
100\,kG. The rotation period is found to be 0.112\,d using the
variable shape of the \ha\ line core, consistent with available
photometry. The field of this star appears to be much more complex than
a dipole, and is probably not axisymmetric.

Available photometry shows that \mwdb\ is a light
variable with an amplitude of only 0.005\,mag; our own photometry shows
that if \mwda\ is photometrically variable, the amplitude is below
about 0.01\,mag.

These are the first models for magnetic white dwarfs with fields below
about 100\,kG based on magnetic measurements through the full stellar
rotation. They reveal two very different magnetic surface
configurations, and that, contrary to simple ohmic decay theory, \mwdb\
has a much more complex surface field than the much younger \mwda. }

\maketitle
%

\section{Introduction}
It has been known since the 1970s that a few percent of white dwarfs
(WDs) have detectable surface magnetic fields
\citep[e.g.][]{Kempetal70, AngeLand71,LandAnge71,Kawketal07}. These fields are
detected in some WDs by the presence of broad-band circular (and
sometimes linear) polarisation, and in other stars by the observation
of Zeeman splitting and circular polarisation in spectral lines such
as the Balmer lines \citep[e.g.][]{Aznaetal04,Kepletal13}. The
observed field strengths, based on measuring Zeeman
splitting or polarisation, or on matching the positions of features in
the spectrum, range from tens of kG (1\,Tesla = 10\,kG) to near
1000\,MG \citep[e.g.][]{Jord92,PutnJord95,Kueletal09,Landetal12} 

Using these methods, during the first thirty years after the discovery
of the first magnetic fields in white dwarfs, new magnetic white dwarfs (MWDs) were discovered
at a rate of about two
per year. The situation changed qualitatively as a result of the Sloan
Digital Sky Survey (SDSS), which has led to the identification of
several hundred new MWDs. However, these two data sets have quite
different properties. The early MWD discoveries mostly are found among
the ``bright'' WDs, with magnitudes of $13 < V < 16$, and many have
been observed at reasonably high \snr, often by
spectropolarimetry \citep[e.g.][]{SchmSmit95}. The fields found in
such stars cover the full range of observed strengths, from a few tens
of kG to about 1000\,MG. In contrast, almost all the SDSS MWDs are in
the range $16 < V < 20$, and the spectra available for these stars
typically have $\snr
\sim 15$ or 20. Due to the low resolving power and low \snr\ values, 
the SDSS data permit the discovery of fields only above a threshold of
1--2\,MG, and field strengths can only be usefully estimated for the
higher \snr\ spectra. Thus the huge increase in the number of
identified MWDs has increased knowledge only of the high-field region
of the full MWD field strength distribution, and most of these new MWDs
can only be studied further using the largest telescopes.

In spite of the large sample of MWDs accumulated through almost 50
years, two fundamental questions remain unanswered. 
\begin{itemize}
\item We do not understand why strong magnetic fields are present in 
a small fraction of white dwarfs, but no detectable fields are found
in the large majority. 
\item We do not yet know how the observed fields evolve as the host 
stars age.
\end{itemize}

In the absence of well-developed theoretical responses to these two
questions, observations can contribute guidance and clues of two
different kinds. First, studies of magnetic characteristics of large
samples of MWDs should make it possible to determine observationally
how the magnitudes of surface fields evolve with WD age, potentially
as a function of WD mass, composition, and space velocity (or
population). Such statistical studies might help to identify source
populations of red giants and/or AGB stars for MWDs of various kinds
\citep[e.g. ][]{Valy15}.  Secondly, detailed studies of individual 
MWDs should provide complementary clues to the ones from statistical
studies. An important kind of information that can emerge from study
of individual stars is the determination of the surface field structure, and
following individual studies, how the structures found depend on field
strength, cooling age, etc.

We are carrying out a large survey, together with detailed studies of
indvidual MWDs, to address several aspects of the two basic questions
listed above. One aspect in which we are particularly interested is
the modelling of the surface magnetic field of the individual MWDs
with the very weakest fields, to obtain empirical information on
whether the field structures are the same as in MWDs with strong
fields, and eventually also how these surface field structures evolve
with WD age. In some strong-field MWDs the observed field is found to vary 
periodically with periods ranging from minutes to weeks
\citep[e.g.][]{Kawketal07,Brinetal13,Ferretal15}, revealing the
rotation period (and total angular momentum) of the underlying
star. This situation allows much more reliable magnetic mapping than
is possible if the observed field is measured only once or twice, or
is unvarying, as the variations make available a variety of views of
the field structure as the star rotates. Accordingly, we are
attempting to determine rotation periods of the super-weak field MWDs,
following up where possible with detailed modelling.






At present, no reasonably constrained surface field model is available
for any MWD with a global field below about 300\,kG. In this paper we
report discovery and analysis of magnetic variability of \mwda\ and
\mwdb, two stars whose fields are among the very weakest clearly
detected WD magnetic fields. We then use the observed variations of
various kinds of measurements of each star to obtain the first
well-constrained surface field models for two MWDs with fields below
100\,kG.


\section{Our targets}
The main physical parameters of the two MWDs discussed in this paper
are given in Table~\ref{Tab_mwd_params}. In the following we provide 
some further details about them.
\begin{table}[th]
\begin{center}
  \caption{Physical parameters of white dwarfs discussed in this paper.
    }
\label{Tab_mwd_params}
\begin{tabular}{lll}\hline
Parameter (units)      &    \mwda         &    \mwdb    \\
\hline
$m_V$ (mag)              & 12.93            & 13.05 \\
$\pi$ (mas)            & $57.87 \pm 0.69$ & $127.4 \pm 6.8$ \\
$D$ (pc)               & $17.3 \pm 0.7$   & $7.85 \pm 0.40$ \\
\te\ (K)               & $14712 \pm 286$  & $8648 \pm 123$ \\
$\log(L/L_\odot)$      & $-2.34$          & $-3.26$ \\
$\log g$ (cm\,s$^{-2}$) & $8.31 \pm 0.04$ & $8.29 \pm 0.05$ \\
$M/M_\odot$            & $0.81 \pm 0.03$  & $0.78 \pm 0.03$ \\
Composition            & H (DA)           & H (DA) \\
Age (Gyr)              & 0.34             & 1.37 \\
\hline
\end{tabular}
\tablefoot{Data taken from \citet{Giametal12} }
\end{center}
\end{table}

\subsection{\mwda}
The weak magnetic field of the DA3.4 \mwda\ = G\,210--36 was first
discovered in October 2015 on the basis of a spectropolarimetric
observation with the intermediate-resolution ISIS spectro-polarimeter on the William Herschel Telescope (WHT), followed by a spectropolarimetric
observation obtained with high-resolution spectropolarimeter \esp on the Canada-France-Hawaii Telescope (CFHT). Both observations revealed clear
splitting of the deep and sharp non-LTE core of \ha, and both provided
marginal detection of Zeeman polarisation in the line wings of \ha. In
addition, weak magnetic splitting was observed in the \esp\ spectrum
of the core of the H$\beta$ line \citep{Landetal16}.

The initial discovery spectra provided measurements of the mean field
modulus (or mean surface field) averaged over the visible hemisphere
\bs\ of about 57\,kG, and values for the mean longitudinal field
averaged over the visible hemisphere \bz\ of about 1\,kG (ISIS) and
6\,kG (\esp). It is worth noting that this WD is quite bright, and
that its spectrum has repeatedly been studied in some detail
\citep[e.g.][]{GreeLieb90,Gianetal11,Giametal12}. A sensitive
low-resolution polarimetric study by \citet{SchmSmit95} did not detect
the magnetic field. In the end, the super-weak field (the
third-smallest field securely detected in a WD) was discovered
using a combination of particularly high \snr, high spectral resolution,
and polarimetry.

\subsection{\mwdb}
The presence of a magnetic field in the bright, cool DA\,5.8 \mwdb\ =
LAWD\,96 was first suggested by \citet{Koesetal98} from the unusually
sharp and shallow core of $H\alpha$ in a high-resolution spectrum of
the star obtained with the European Southern Observatory (ESO) high-resolution UVES spectrograph on the Very Large Telescope (VLT) for the
supernova progenitor (SPY) programme. The presence of a field was
confirmed using low-resolution spectropolarimetry with the ESO FORS1 low-resolution
spectropolarimeter by \citet{Aznaetal04}, who detected fields of $\bz \approx
+3$ and $+4$\,kG at the $5-6\sigma$ level of significance on two
different nights. (Note that these data were re-reduced as described
by \citet{Bagnetal15}, and the sign of the observations changed to
conform to the usual convention.) The value of \bs\ was subsequently
estimated from the UVES spectra to be about 110\,kG
\citep{Koesetal09}. Like \mwda, this bright WD had been studied
extensively but was only recognised as magnetic from high-resolution
spectroscopy.

\begin{table*}
  \caption{\label{Tab_All_Meas} Summary of the measurements used in this work.
} 
\begin{tabular}{lll}
\hline\hline\\
                     & \mwda                                         & \mwdb                          \\
\hline
                     &                                               &                                \\
    HR Spectroscopy  &                                               & 2 UVES observations \citep{Koesetal09}   \\[2mm]
LR Spectropolarimetry&                                               & 2 FORS1 observations \citep{Aznaetal04}  \\
                     &                                               & 4 FORS2 observations (this work)      \\[2mm]
MR Spectropolarimetry& 1 ISIS observations \citep{Landetal16}        &                                \\
                     & 4 MSS observations (this work)                &                                \\[2mm]
HR Spectropolarimetry& 1 ESPaDOnS observation \citep{Landetal16}     & 12 ESPaDOnS observations (this work)     \\
                     & 16 ESPaDOnS observations (this work)          &                                \\[2mm]
    Photometry       & SAO RAS time series (this work)            & Amateur astronomer time series \citep{Garyetal13} \\
    \hline
  \end{tabular}
\tablefoot{LR, MR, HR stand for Low-, Mid-, High-resolution,
  respectively.}
\end{table*}

\section{Observations}

The data used in this work are summarised in Table~\ref{Tab_All_Meas}
and come from different sources. Some datasets come from our own
observing programmes and are presented here for the first time, and
some have been taken from ESO archive and/or have been already
published (either by ourselves or by other groups). 

Our general observing plan was first to identify a candidate weak
field magnetic star (either from the literature or using data from our
surveys), then to try to detect variability either in intensity or in
polarisation spectra with a small number of observations per star;
finally, when variations were found, to monitor the star with a
cadence suitable to detect a period in the range of hours to weeks.
This required a series of observations spaced one or a few days apart,
combined with either some repeated observations per night (\mwda) or
at least observations at a range of hour angles (\mwdb). \esp\ at the
CHFT (see Sect.~\ref{Sect_ESP} below) was the principal instrument
used for this longer term monitoring.  Photometric data were also used
to confirm/identify a rotation period.




\subsection{Mid-resolution spectropolarimetry with MSS of the SAO 6\,m telescope}

The MSS is a moderate-beam classic spectrograph equipped with a
circular polarization analyser combined with a 7-layer image slicer
\citep{Chou04} permanently installed at the 6-m telescope (BTA) at the
Special Astrophysical Observatory (SAO) of the Russian Academy of Sciences (RAS).
The description of the optical scheme of the MSS and its capabilities
can be found in the web page of the instrument\footnote{\tt
  http://www.sao.ru/hq/lizm/mss/en/ }. The rather dated design of the
instrument, which has several mirrors along the beam path, seriously
reduces total efficiency in high-resolution observations of
comparatively faint white dwarfs. To respond to this problem the
observatory staff is currently constructing a modern, high-resolution
spectropolarimeter which is expected to be commissioned within a
couple of years; see \citet{Valyetal14B}. With this instrument it will
be possible to extend BTA observations of white dwarfs up to $m_V \sim 15$
or fainter. Our current observational limit with the
high-resolution spectropolarimeters of the 6-m telescope is around $V
\sim 13$. Nevertheless we decided to use this instrument in order to
obtain as much information about \mwda\ as possible, for which we
obtained four spectropolarimetric observations in the course of two
consecutive nights, 2016-06-12 and 13. The MSS spectropolarimeter was
used with spectral resolving power of $R=10\,000$ at \ha.  The
observations were conducted following the standard procedures that we
use in observations with BTA \citep[see for example][]{Valyetal05}. A
polarisation observation consists of a series of paired exposures
obtained at two orthogonal orientations of the quarter
wave-plate. Data reduction and field measurements are also standard
and summarised by \citet{Landetal15}.
The log of MSS observations is given in Table~\ref{Tab_meas_2047}.

\subsection{Low-resolution spectropolarimetry with FORS of the ESO VLT}\label{Sect_FORS}

FORS1 and FORS2 \citep{Appetal98} are twin multipurpose instruments
capable of doing imaging and low-resolution spectroscopy attached at
the cassegrain focus of the 8\,m units of the ESO VLT. When both
instruments were operating, FORS1 was equipped with polarimetric
optics. Polarimetric optics were then moved to FORS2 when FORS1 was
decommissioned in 2008. Both instruments have been used for
observations of WDs by several authors. In this paper we consider UVES
and FORS1 archive data of \mwdb\ as well as four new observations
obtained during our survey. Our new spectra cover the wavelength
window between about 3670 and 5120\,\AA\ and have an effective
resolving power of about 2200. They have a typical S/N ratio per pixel
of between 270 and 400 per \AA. These data were reduced, and the mean
longitudinal field \bz\ deduced, as described in detail by
\citet{Bagnetal15}: for these spectra \bz\ was determined by measuring
the slope of the correlation between the circular polarisation
$V(\lambda)$ and $(1/I) ({\rm d}I(\lambda)/{\rm d} \lambda)$ (see
Sect.~\ref{Sect_Bz}).  The log of FORS2 observations is given in
Table~\ref{Tab_meas_2359}.

\subsection{High-resolution spectropolarimetry with ESPaDOnS of the CHFT}\label{Sect_ESP}

In order to study the magnetic fields of the two super-weak field
MWDs investigated in this paper in more detail, we obtained observations with the high-resolution
CFHT spectropolarimeter \esp, which we had previously found to be a
very powerful tool for study of fields in bright DA WDs
\citep{Landetal15,Landetal16}. The instrument provides high-precision,
high-resolution spectropolarimetry over nearly the complete
spectrum from 3800\,\AA\ to 1.04\,$\mu$m using a polarimetric module
above the entrance apertures of a cross-dispersed echelle
spectrograph, as briefly described by the instrument web
page\footnote{\tt
  http://www.cfht.hawaii.edu/Instruments/Spectroscopy/Espadons/} and
by \citet{Landetal08}. \esp\ has very high throughput, but with
$R=65\,000$ on a 3.6-m telescope, observing even relatively bright WDs
is uncomfortably near the magnitude limit set by cosmic rays, CCD read
noise, sky brightness near twilight or when the moon is near full, and
the need to keep each observation from lasting too long in order to
sample a reasonable range of probable rotation periods. As a
compromise, each observation (comprising four sub-exposures taken at
different wave-plate positions) lasted about an hour in total. This
integration time was found (after binning spectra to $R \approx
15\,000$) to provide sufficient \snr\ to clearly see magnetic effects
in the deep and sharp non-LTE cores of the \ha\ lines of the two MWDs
with field uncertainties of the order of one or a few kG
\citep{Landetal15,Landetal16}.

The CFHT allocated time during the 2016A and 2016B semesters, and with the
flexibility and excellent consistency provided by the queue service
observing team, we were able to get two series of polarised
spectra. We note that the series of \esp\ spectra obtained for \mwdb\ is
particularly remarkable, because the star's position at $-43^\circ$
guaranteed that all spectra were taken at more than two air masses. In
spite of this, nine spectra were obtained during ten nights, over a
sufficient range in hour angle to make possible the secure discovery
of periodic magnetic variations with a period of about 2.6\,hr (see below).
The log of \esp\ observations is given in Tables~\ref{Tab_meas_2047}
and \ref{Tab_meas_2359}.
\begin{figure*}[ht]
  \begin{center}
  \scalebox{0.36}{\includegraphics*[trim={0.8cm 1.0cm 2.3cm 2.8cm},clip]{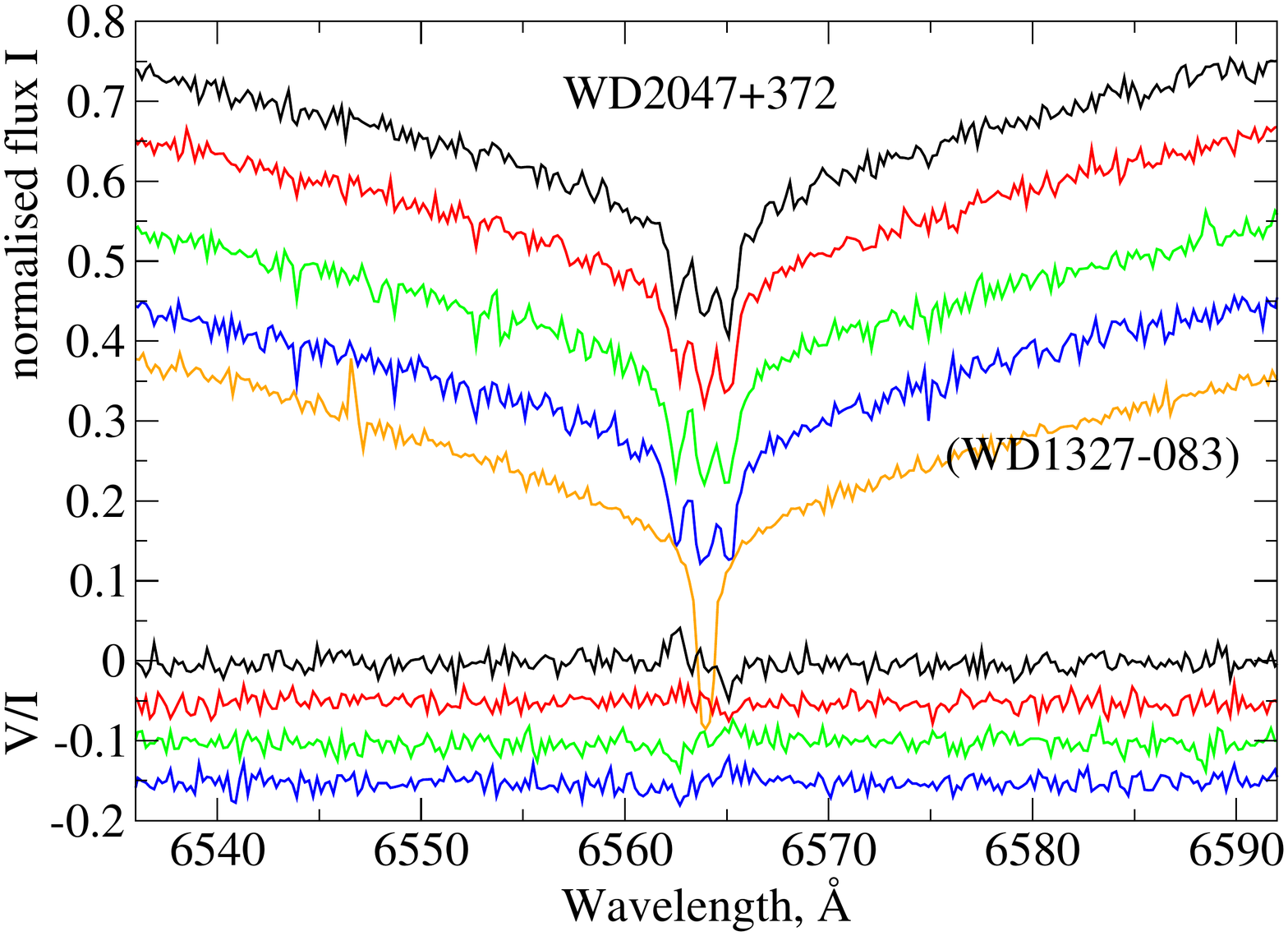}}
  \scalebox{0.36}{\includegraphics*[trim={0.8cm 1.0cm 2.3cm 2.8cm},clip]{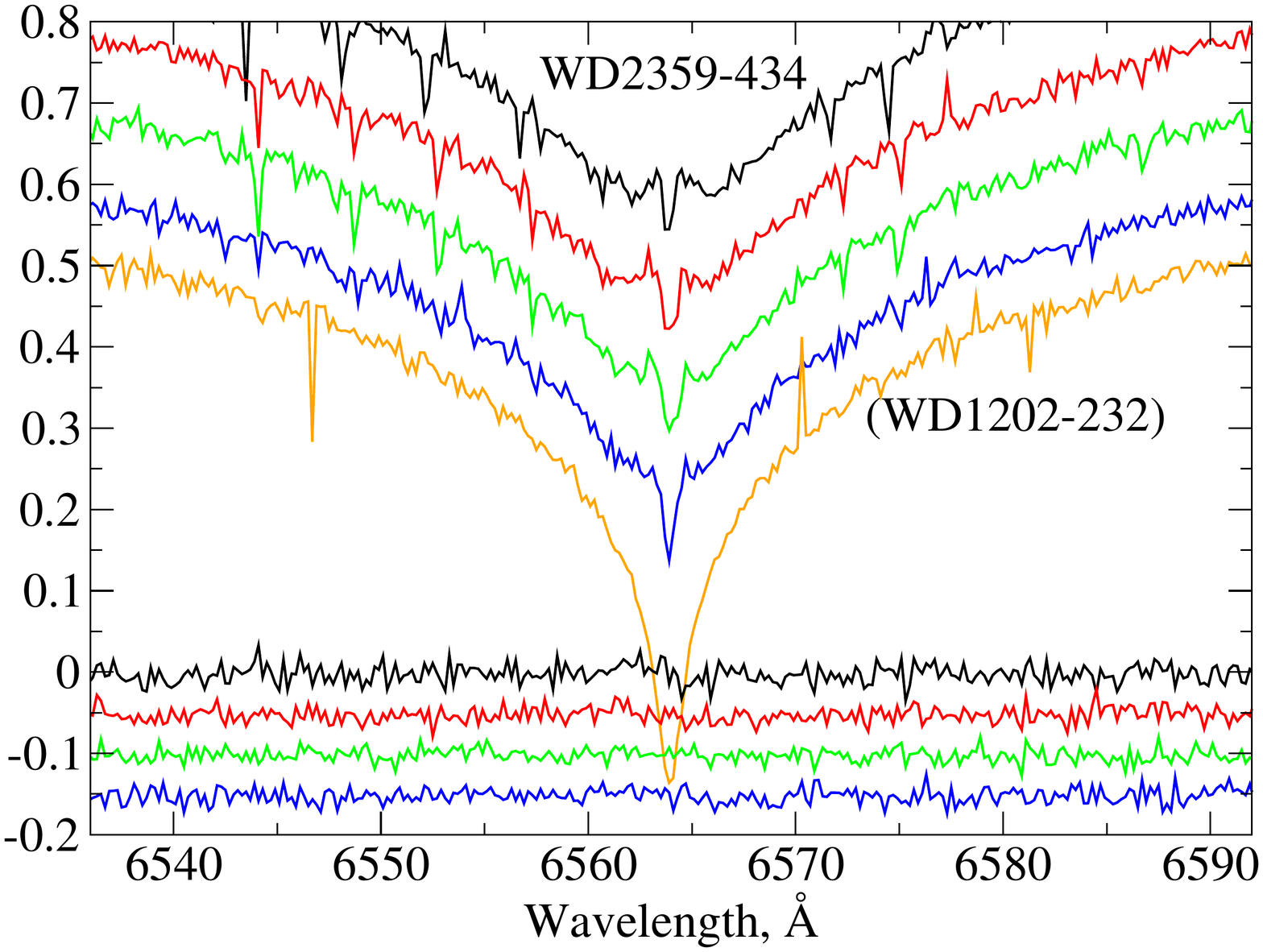}}
\end{center}
  \caption{\label{Fig_esp}
\esp\ spectra of \mwda\ (left panel) and \mwdb\ (right panel). Both
panels are organised in a similar way. The top four solid lines show
the intensity $I$ spectra, and the lowest solid lines the
corresponding circular polarisation $V$ spectra of the \ha\
region. For \mwda\ the spectra numbers are 1975651, 1976104, 1976387,
1976638; for \mwdb\ the spectra are 1975659, 1976391, 1976646, and
1974336.  In between the four \esp\ $I$ spectra is a UVES $I$ spectrum of a
non-magnetic star with similar effective temperature as the magnetic
star: the comparison star for \mwda\ is WD1327--083, and the
comparison star for \mwdb\ is WD1202-232. Data are binned in
0.2\,\AA\ boxes to improve \snr.  All $I$ spectra are normalised to
1.0 at $\pm 200$\,\AA\ from \ha\ line centre, then shifted down from
top $I$ spectrum in steps of 0.1 for visibility. All $V$ spectra are
normalised to 0.0 outside the line core, then shifted down from top
$V$ spectrum in steps of 0.05.
}
\end{figure*}

\subsection{Photometry with the SAO 1\,m telescope }
\begin{table}
 \caption{Journal of photometric observations of \mwda\ with the 1-m
telescope of the SAO--RAS.}
 \label{Tab_sao_phot}
 \begin{center}
  \begin{tabular}{llr}
\hline
\hline
Date & JD (days) & $N_{\rm exp}$ \\
\hline
2016-08-19 & 2457620.31 -- 20.57 & 1083 \\
2016-09-12 & 2457644.33 -- 44.39 & 238  \\
2016-09-19 & 2457651.22 -- 51.41 & 513  \\
\hline
\end{tabular}
\end{center}
\end{table}

In order to supplement our magnetic mesurements we have also conducted
photometric observations of \mwda. These observations were obtained
during three nights (August 19, September 12 and September 19, 2016) with the
1-m telescope of the Special Astrophysical Observatory of the Russian
Academy of Science (see Table~\ref{Tab_sao_phot}). All the observations were conducted in the
broadband $m_V$ filter of the Johnson system as a continuous run of
short duration individual exposures, typically 20--30\,s in
length. Before each night of observations, exposures of scattered
sunlight were made during twilight for flat field correction. The
photometric data reduction was carried out with software written in
the Python programming language using the standard package for
astrophysical data reduction from the Image Reduction and Analysis Facility (IRAF) of the U.\,S.\,National Optical Astronomy Observatories.

\subsection{Amateur astronomer photometry}

Broad-band photometric monitoring of \mwdb\ has been undertaken by a
group of amateur astronomers, particularly by T.\ G.  Tan in Perth,
Australia, using telescopes
of around 30\,cm aperture. This work was initiated as an effort to
detect planetary transits in white dwarf systems. When sinusoidal light
variations of \mwdb\ were found, these were followed up. A major
observing campaign was carried out on this star during 2011 and 2012,
involving many tens of hours of observing  \citep{Garyetal13}. Further data were acquired
during 2013, 2014, and 2016\footnote{\tt
http://brucegary.net/WDE/WD2359-434/WD2359-434.htm}. Most of the observations were carried out
in either the $Rc$ or $B$ photometric bands. The precision of
individual photometry measurements on good nights, with integrations
times of order 1\,min, is in the vicinity of 3--5\,mmag.

\section{Measurements of spectropolarimetric data}
Our spectral database includes high-resolution \esp\ polarised spectra
for both stars. These spectra can be conveniently used for a preliminary
quantitative comparison of the magnetic fields of the two stars.

Most of the useful information in the \esp\ spectra about the fields
of the two MWDs is found within a few \AA\ of the centre of \ha. Four
representative flux spectra (Stokes $I$ spectra) of \mwda\ in this spectral region are shown in
the left panel of Fig.~\ref{Fig_esp}, together with a comparison
spectrum of the non-magnetic and unpolarised star WD1327-083 ($\te =
14570$\,K), and four typical spectra of \mwdb\ are shown in the right
panel of Fig.~\ref{Fig_esp}, together with a UVES comparison spectrum
of the non-magnetic star WD1202-232 ($\te = 8770$\,K).

In both stars, clear (but quite different) Zeeman splitting is visible
in the intensity spectra. In \mwda\ the Zeeman pattern contains sharp
$\sigma$ components, very similar in shape to the central $\pi$
components. The shape of the clear and simple Zeeman flux triplet
pattern hardly varies from one observation to another. Significantly non-zero 
Zeeman polarisation signals, varying from one spectrum to another, are
visible to the eye in most of the Stokes $V$ (circular polarisation) spectra
of \mwda. In contrast, only the central $\pi$ component of the Zeeman
split line core is obvious in \mwdb, while the two $\sigma$ components
are broad and very poorly defined. The entire shape of the line core
of this star is variable. In the \esp\ $V$ spectra of \mwdb, non-zero 
Zeeman polarisation signals are barely detectable to the eye, although non-zero
values of \bz\ are securely detected at the several sigma level in all
six FORS polarised spectra.

The \esp\ spectra were used to extract the following information:
\begin{itemize}
\item[] Variability of the \ha\ line core profile.
\item[] The equivalent width of \ha.
\item[] The mean longitudinal magnetic field \bz.
\item[] The mean magnetic field modulus \bs.
\end{itemize}
FORS low-resolution, and ISIS and MSS mid-resolution spectra, were
used only to measure the mean longitudinal magnetic field.

In the following we describe these measurements, and we report the
results obtained for the two stars. Our measurements will be used
below (Secs. 5 and 6) to confirm stellar variability, search for its
period, and finally to obtain approximate magnetic models of our
target stars.

\subsection{Line core variability}
The high resolution \esp\ spectra may be used to test whether the
\ha\ line core is detectably variable or not as follows. We first
construct a mean inner \ha\ line by averaging all available
spectra. This averaging is carried out on the normalised spectra
provided by the CFHT's Libre-Esprit reduction pipeline, which produces
spectra that are consistently normalised in a very similar
way. Individual spectra are then slightly rescaled to coincide as well
as possible with the mean spectrum. After rescaling, the standard
deviation $\sigma_{\rm csp}(\lambda_j)$ of the ensemble of spectra is
calculated point by point as a function of $\lambda_j$ through the
profile using
\begin{equation}
\sigma_{\rm sp}(\lambda_j) = \left\{\frac{\sum_{i=1}^{n} 
[y_i(\lambda_j) - y_{\rm m}(\lambda_j)]^2}{n-1}\right\}^{1/2}, 
\end{equation}
where $i$ is summed over the $n = 16$ spectra at each point in the
spectrum, and $y_{\rm m}(\lambda_j)$ is the value of the mean spectrum
at that wavelength. 

\begin{figure*}[ht]
  \begin{center}
\scalebox{0.36}{\includegraphics*[trim={0.8cm 1.0cm 2.3cm 2.8cm},clip]{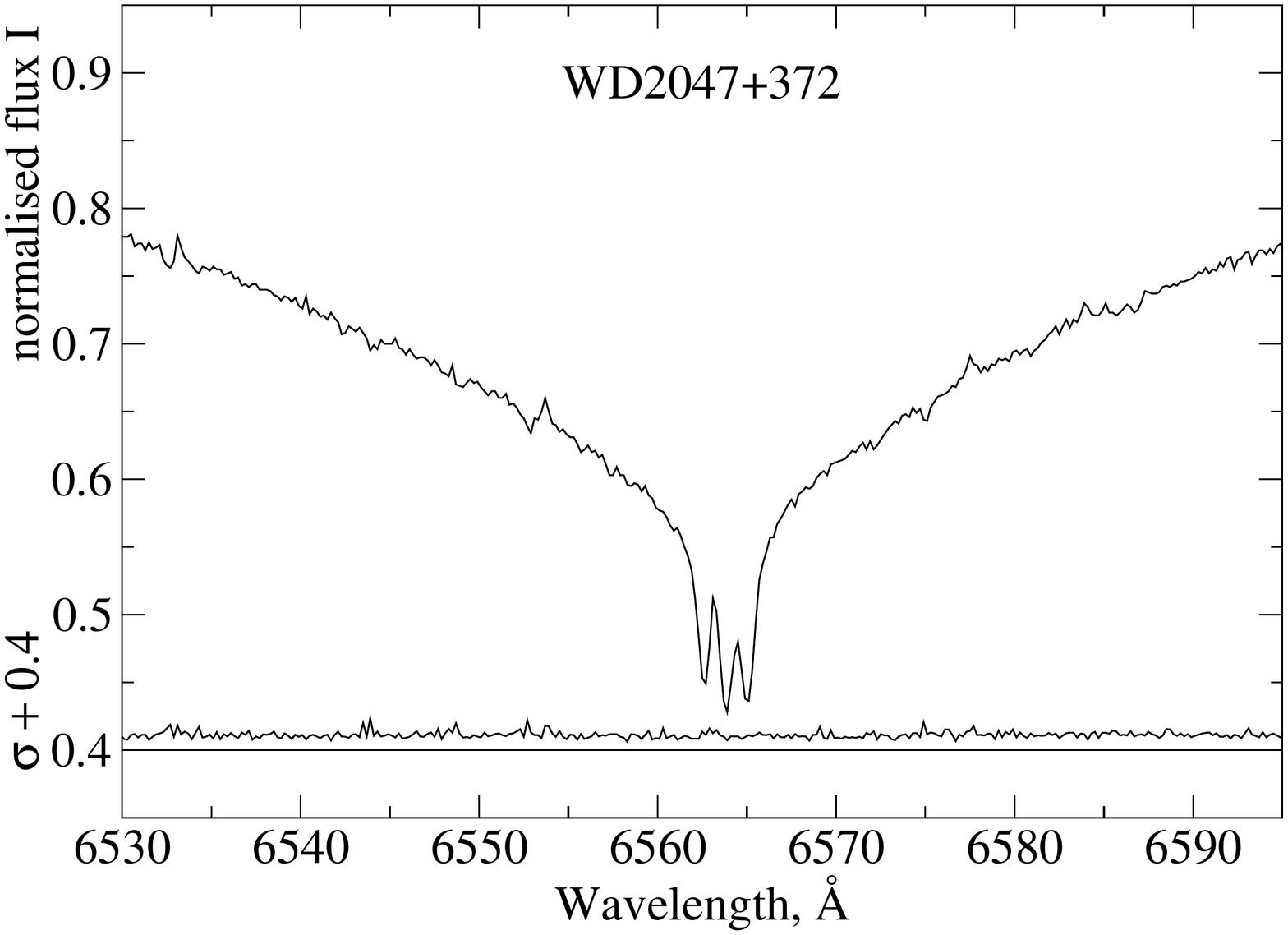}}
\scalebox{0.36}{\includegraphics*[trim={0.8cm 1.0cm 2.3cm 2.8cm},clip]{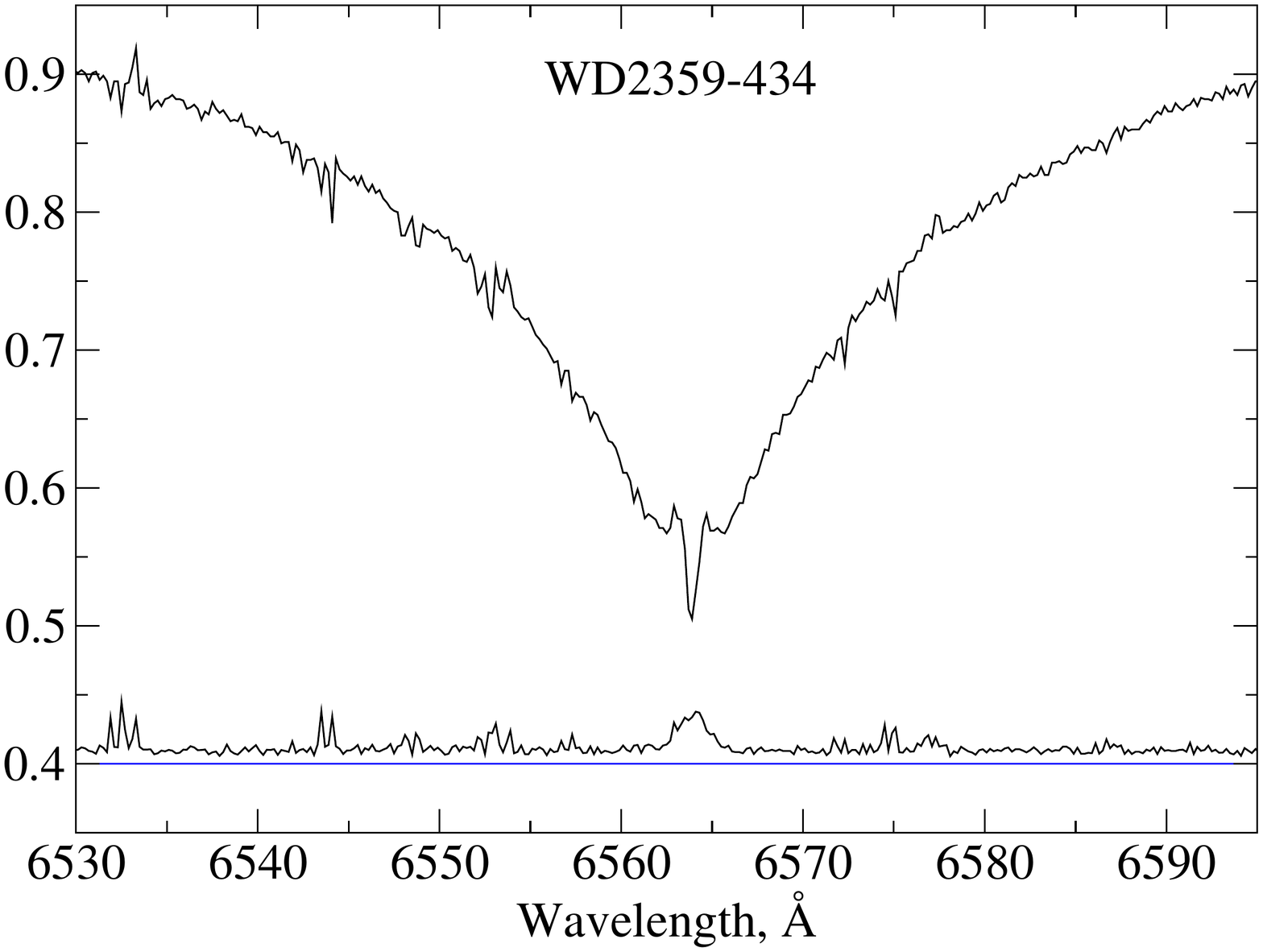}}
\end{center}
\caption{\label{Fig_esp_mean_spec_sigma} Mean \ha\ flux ($I$)
spectrum computed from all individual spectra), with standard deviation $\sigma$ of individual spectra
around the mean, shifted by +0.4 for clarity. The straight line at $+0.4$
is the zero point for $\sigma$, also shifted upward by $+0.4$. Left panel
refers to \mwda\ (excluding spectrum E 1939614) and right panel
refers to \mwdb.}
\end{figure*}

This mean spectrum of \mwda\ is shown in the left panel of
Fig.~\ref{Fig_esp_mean_spec_sigma}. The variation of
$\sigma_{\rm sp}(\lambda_j)$ across the core of \ha\ is shown below
the mean $I$ line profile (shifted upward by +0.4 from its original
value close to zero, for ease of comparison with the $I$ profile; the
zero line has also been shifted upward by +0.4 and is seen about 0.015
below the computed standard deviation).  The striking feature of the
variation of $\sigma_{\rm sp}$ across the line core where the Zeeman
triplet is found is that there is {\em no evidence for larger
  dispersion}. This confirms what is anticipated above, i.e., that the
available \ha\ spectra show no evidence that the value of \bs\ varies
significantly as the WD rotates; \bs\ is constant at about 60\,kG.

In contrast to \mwda, \mwdb\ (right panel of
Fig.~\ref{Fig_esp_mean_spec_sigma}) exhibits a clear increase in
dispersion across the core of \ha\ (and also across several of the
weak atmospheric water vapour absorption lines present in the stellar
line wings). This variability is also visible in the $I$ spectra of
Fig.~\ref{Fig_esp}.

\subsection{Equivalent width}
The simplest moment of the \ha\ $I$ profile is the equivalent width of
the Zeeman-split core. In the context of this work, EW measurements
serve to detect periodic variability. They were found useful only in
the case of \mwdb, for which they are also found to act as proxies for
\bs. To obtain EW measurements, we first renormalised each \esp\ \ha\ 
profile (as produced in normalised form by the CFHT reduction
programme Libre-Esprit) to minimise the mean squared difference from
the mean of all $I$ \ha\ spectra, using the two wing intervals $6530 -
6560$\,\AA\ and $6568 - 6598$\,\AA\ (omitting small intervals with
significantly variable atmospheric absorption). This led only to small
amounts of rescaling, with changes of 2\,\% or less. We then
integrated the core area between 6560\,\AA\ and 6567.8\,\AA\ relative
to an arbitrary ``continuum'' level at 0.6. The resulting equivalent
width measures $W_{\lambda}$, in m\AA, are reported in
Table~\ref{Tab_meas_2359}. Note that although the values have an
arbitrary zero point (unlike the situation for a sharp line isolated
in a local continuum), the {\em variations} have the same meaning as
usual for equivalent width variations. We estimate that the total
uncertainty of each of these measurements is roughly 20\,m\AA, mainly
due to uncertainty in the slight renormalisation of the spectra to a
common mean.

\subsection{The mean longitudinal magnetic field}\label{Sect_Bz}

\begin{table}
\begin{center}
  \caption{\label{Tab_meas_2047}
    Journal of observations of \mwda.}
\begin{small}
  \begin{tabular}{llrr@{$\pm$}lr@{$\pm$}l}
  \hline\hline
Instrument &  MJD      &Exp.&\multicolumn{2}{c}{\bz}&\multicolumn{2}{c}{\bs} \\
           &           &(s)&\multicolumn{2}{c}{(kG)} &\multicolumn{2}{c}{(kG)}\\
\hline
 ISIS      & 57266.082 & 3360 & $   0.61 $& 0.24 \\
E 1844823  & 57326.255 & 3256 & $   6.00 $& 1.67 & 57.4 & 2.5\\
E 1939267  & 57549.554 & 4416 & $ -10.79 $& 1.37 & 59.8 & 2.5\\
E 1939614  & 57551.601 & 4416 & $  16.11 $& 2.77 & 61.3 & 3.0\\
MSS        & 57551.942 & 3600 & $   2.20 $& 2.00 \\
MSS        & 57551.984 & 3660 & $  -2.80 $& 2.00 \\
MSS        & 57552.901 & 3660 & $  -9.70 $& 1.80 \\
MSS        & 57552.944 & 3660 & $  -7.60 $& 2.30 \\
E 1940930  & 57558.609 & 4416 & $  -1.82 $& 1.13 & 55.3 & 2.5\\
E 1941326  & 57560.511 & 4416 & $ -11.69 $& 1.72 & 59.5 & 2.5\\
E 1941535  & 57561.500 & 4416 & $  -9.30 $& 1.26 & 60.5 & 2.5\\
E 1941724  & 57562.451 & 3400 & $  -9.57 $& 1.60 & 59.5 & 2.5\\
E 1941744  & 57562.611 & 3400 & $   8.97 $& 1.59 & 57.4 & 2.5\\
E 1974328  & 57603.483 & 3360 & $   3.25 $& 1.31 & 58.9 & 2.5\\
E 1974340  & 57603.577 & 3360 & $  -6.46 $& 1.55 & 63.0 & 2.5\\
E 1974958  & 57606.380 & 3360 & $   6.89 $& 1.46 & 60.4 & 2.5\\ 
E 1974982  & 57606.553 & 3360 & $  13.44 $& 2.03 & 61.9 & 2.5\\ 
E 1975651  & 57609.499 & 3360 & $  14.66 $& 1.51 & 60.2 & 2.5\\ 
E 1976104  & 57611.406 & 3360 & $   6.16 $& 1.23 & 57.9 & 2.5\\
E 1976387  & 57612.556 & 3360 & $ -10.97 $& 1.38 & 61.5 & 2.5\\
E 1976638  & 57613.487 & 3360 & $  -8.65 $& 1.23 & 64.4 & 2.5\\
E 1977043  & 57615.441 & 3360 & $ -10.32 $& 1.34 & 58.8 & 2.5\\
\hline
\end{tabular}
\end{small}
\end{center}
\tablefoot{Column~1 gives the instrument used (\esp\ spectra are
identified by the letter E followed by the identification number);
Col.~2 gives the Modified Julian Date at mid-exposure, Col.~3 the
exposure time, Col.~4 the measured longitudinal field \bz\ with
its uncertainty, and Col.~5 the mean field modulus \bs\ and its
estimated uncertainty.}
\end{table}
\begin{table}[th]
\begin{center}
  \caption{Journal of observations of \mwdb.}
  \label{Tab_meas_2359}
  \begin{small}
\begin{tabular}{llrr@{$\pm$}lrr}
  \hline\hline
Instrument &  MJD        & Exp.   &\multicolumn{2}{c}{\bz} &\bs& $W_{\lambda}$ \\
           &             & (s)    &\multicolumn{2}{c}{(kG)} &kG &  (m\AA)     \\
\hline
UVES       &  51739.276  &   300  & \multicolumn{2}{c}{}& 110  &         \\
UVES       &  51743.301  &   300  & \multicolumn{2}{c}{}& 110  &         \\
FORS1      &  52583.025  &  2188  & $ 4.10 $& 0.84 &      &$       $\\  
FORS1      &  52608.056  &  2188  & $ 3.09 $& 0.51 &      &$       $\\
FORS2      &  57234.299  &  4296  & $ 2.14 $& 0.37 &      &$       $\\
FORS2      &  57273.252  &  4296  & $ 2.94 $& 0.42 &      &$       $\\
E 1844491  &  57324.324  &  3256  & $10.81 $& 3.28 &  99  &$  38.4 $\\ 
E 1844835  &  57326.343  &  3256  & $ 3.50 $& 2.86 &  99  &$  17.5 $\\ 
E 1844989  &  57327.289  &  3256  & $ 4.80 $& 1.76 &  57  &$ 304.7 $\\
E 1845170  &  57328.285  &  3256  & $ 5.70 $& 2.15 &  74  &$ 212.8 $\\
FORS2      &  57544.374  &  2496  & $ 2.40 $& 0.40 &      &$       $\\  
FORS2      &  57567.399  &  2496  & $ 2.10 $& 0.34 &      &$       $\\  
E 1974336  &  57603.533  &  3360  & $ 6.69 $& 1.80 &  47  &$ 347.7 $\\ 
E 1974574  &  57604.542  &  3360  & $-0.59 $& 2.03 &  49  &$ 317.6 $\\ 
E 1974986  &  57606.598  &  3360  & $ 5.43 $& 2.59 &  76  &$ 152.9 $\\ 
E 1975449  &  57608.493  &  3360  & $ 8.37 $& 2.06 &  61  &$ 234.6 $\\ 
E 1975659  &  57609.546  &  3360  & $ 8.57 $& 3.46 & 100  &$ -12.3 $\\ 
E 1975887  &  57610.585  &  3360  & $ 8.89 $& 2.41 &  71  &$ 164.8 $\\ 
E 1976112  &  57611.476  &  3360  & $ 5.69 $& 2.07 &  73  &$ 150.9 $\\ 
E 1976391  &  57612.599  &  3360  & $ 7.49 $& 2.51 &  81  &$ 113.2 $\\
E 1976646  &  57613.556  &  3360  & $ 6.23 $& 2.50 &  67  &$ 212.2 $\\
\hline
\end{tabular}
\end{small}
\end{center}
\tablefoot{Column~1 gives the instrument used (\esp\ data are
  identifed by the letter E followed by the identification number),
  Col.~2 the Modified Julian Date at mid-exposure, Col.~3 the exposure
  time, Col.~4 the measured longitudinal field \bz\ with its
  uncertainty, and Col.~5 the mean field modulus in kG (which has a
  typical estimated uncertainty of 10\,kG), and Col.~6 the measured
  \ha\ line core EW $W_{\lambda}$. \bs\ values for UVES spectra are
  from \citet{Koesetal09}.}
\end{table}

The mean longitudinal magnetic field can be estimated in two different
ways. From mid- and low-resolution spectra we have used the
relationship between $V(\lambda/I)$ and $(1/I) ({\rm d}I(\lambda)/{\rm
  d} \lambda)$ \cite[e.g.][]{Bagnetal15}. From high-resolution
\esp\ spectra of \ha, the mean longitudinal field can conveniently be
derived by measuring the separation between the wavelength of the
centroid of the line core profile as seen in right polarised light and
the wavelength of the centroid of the line core in left circularly
polarised light \citep{Math89,Donaetal97}.  \bz\ was computed
separately for the \ha\ and H$\beta$ lines. For \mwda\ the integration
limits were 6561.5 and 6566.0\,\AA\ for \ha, and 4860.0 and
4864.2\,\AA\ for H$\beta$. The two \bz\ values were then combined by
weighted averaging. Note that in general the value of \bz\ from
H$\beta$ has an uncertainty about three times larger than the value
derived from \ha, so the contribution of H$\beta$ to the final values
is almost insignificant. Details of the numerical application of this
method to WD \esp\ spectra are discussed by \citet{Landetal15}.


The values of \bz\ derived from each of the spectra of \mwda\ are
reported in the third column of Tab.~\ref{Tab_meas_2047} \cite[note
  that the ISIS measurement and the first \esp\ measurement were
  already published by][]{Landetal16}.  In 15 of 17 measurements obtained
with ESPaDOnS, the field is detected at more than the
$3\,\sigma_{Bz}$ level. The values of the uncertainty $\sigma_{Bz}$ are
generally between 1 and 2\,kG, so the value of \bz\ varies by
several times $\sigma_{Bz}$ on each side of zero. Thus these data are
well-adapted to searching for time variability. We note that the ISIS
measurement has far higher precision than the \esp\ measurements.

For \mwdb\ the integration limits to use in determination of \bz\ are
not obvious; the sigma components blend very smoothly into the broader
line wings. For this star, the integration limits were chosen to be
just a little beyond the edges of the apparent region of spectral
variation (and the limit of non-zero circular polarisation in an
averaged spectrum computed for several spectra obtained around maximum
field). The limits chosen were 4859.5  and 4864.7\,\AA\ for H$\beta$,
and 6559.0 and 6568.6\,\AA\ for \ha.

The values of \bz\ derived from each of the spectra of \mwdb\ are
reported in Col.~3 of Table~\ref{Tab_meas_2359}. It is clear that the
FORS measurements have much smaller error bars than those obtained
with \esp\ and MSS. The large uncertainties of the \esp\ measurements
relative to the FORS measurements are a consequence of the fact that
for this star we are measuring the small wavelength shift of a single
broad, shallow feature between the right and left circularly polarised
spectra, using only a few \AA\ around the centre of \ha; in contrast,
in the FORS data we are also measuring about the same small shift
using broad line wings, but in the FORS data the full wings of five
Balmer lines (starting with H$\beta$) contribute to the S/N of the
measurement.

\subsection{The mean magnetic field modulus}
Another measurement of direct interest for characterising the surface
magnetic field is the mean field modulus \bs, a quantity representing
the local surface values of $\vert B \vert$, averaged over the visible
hemisphere of the star. This field moment is determined by measuring
the separation of the two outer ($\sigma$) Zeeman components of the
magnetically split line from the central ($\pi$) component, and
deriving the value of \bs\ by using the expression (adapted from the
equations describing the Zeeman effect)
\begin{equation}
  \Delta \lambda_{\rm Z} = 4.67\,10^{-13} \bar{g} \lambda_0^2 \bs\ \,
  \label{Eq_Bs}
\end{equation}
where $\Delta \lambda_{\rm Z}$ is the $\pi - \sigma$ separation,
$\lambda_0$ is the wavelength of the spectral line in the absence of a
magnetic field, \bs\ is the magnetic field strength, and $\bar{g}$ is
the mean Land\'e factor (the mean shift of the Zeeman $\sigma$
components from the unperturbed line wavelength, measured in units of
the clssical Zeeman splitting). For H$\alpha $, $\bar{g} = 1$
\citep{CasiLand94}. In Eq.~(\ref{Eq_Bs}) all wavelengths are in
\AA\ units, and field strength is measured in Gauss (G). Note that we
usually actually measure the separation of the two $\sigma$
components, which is twice the separation described by Eq.~(2). 

In the discovery spectrum of the field of \mwda\ we estimated the
value of $\bs = 56.9 \pm 0.4$\,kG, where the uncertainty was simply an
estimate of the uncertainty of repeatedly measuring the $\sigma -
\pi$ separation wavelengths, using Gaussian fits to the three Zeeman
components. We have measured the value of \bs\ for all the \esp\
spectra using the separation between the two $\sigma$ components,
where the centroid wavelength of each $\sigma$ component was measured
using the {\it e - e} command of the IRAF spectrum plotting and
measuring function {\it splot}. The resulting \bs\ measurements are
reported in Tab.~\ref{Tab_meas_2047}. It is clear on inspection that
the values of \bs\ scatter around a mean of about 60\,kG with about
the scatter expected from our estimates of measurement uncertainty
(based on experiments with different ways of measuring the $\sigma -
\sigma$ separation). Note that the measurement uncertainties reported for \bs\ now effectively include an uncertainty for the transformation from $\sigma - \sigma$ component separation to \bs, and hence are significantly larger than the uncertainty reported in our initial paper on \mwda. There is no convincing evidence in these \bs\
data that \bs\ in this MWD varies significantly. 

The Zeeman patterns of \mwdb\ are remarkably different from those of
\mwda, although both stars appear to have fields of somewhat less than
100\,kG. Where the Zeeman $\sigma$ components of \mwda\ are not
significantly wider than the $\pi$ component, those of \mwdb\ are so
broad that identifying them is difficult. It seems clear that the
local variations in the value of $|B|$ seen on at least part of the
stellar surface reach a factor of two or even more, while on \mwda\ it
is not obvious that $|B|$ varies at all over a visible hemisphere.

As a result, in \mwdb\ it is very difficult to know how to measure the
field moment \bs, as in this star the $\sigma$ components of the
Zeeman triplet are broad and shallow, have unfamiliar shapes, have low
S/N relative to the overall profile, and blend imperceptibly into the
line wings. However, close examination of the individual spectra
strongly suggests that the positions of the Zeeman $\sigma$ components
varies considerably from one spectrum to another, from positions well
clear of the central $\pi$ component to almost merging with the $\pi$
component. Thus it appears that in this star \bs\ is rather strongly
variable.

We experimented with a number of methods of obtaining reproducible and
meaningful measurements of \bs. Reasonable estimates by eye of the
positions of the $\sigma$ components of the line profiles could be
made for some of the spectra, especially those in which the $\sigma$
components are relatively obvious (e.g. the top spectrum in
Fig.~\ref{Fig_esp}). However, this method was often foiled by noise
(even in spectra smoothed to 0.2\,\AA\ bins) because of the very
shallow shape of these components. The resulting measurements were
found to be too noisy to be very useful, although they do suggest that
the value of \bs\ varies between nearly 100\,kG and below about 60\,kG.

Using {\it e-e} command of the IRAF {\it splot} function allowed us to
estimate the position of the centroid of each of the $\sigma$
components of the \ha\ line core in the various spectra, but this
quantity showed only very small variations in the deduced values of
\bs\ (roughly from 100 to 80\,kG) in spite of the obvious
variations of the line shapes. We concluded that this particular
measure is not very sensitive to the value of \bs.

Table~\ref{Tab_meas_2359} shows that one measured quantity, our measure
of the equivalent width $W_\lambda$ of the line core in Stokes $I$,
varies from spectrum to spectrum by much more than the associated
uncertainty. We therefore considered using this quantity as a proxy
for \bs. Inspection of individual spectra clearly shows that
$W_\lambda$ is closely associated with the form of the $\sigma$
components. In spectra in which the $\sigma$ components are relatively
clear and well separated from the $\pi$ component, the line core is
broad and shallow, and $W_\lambda$ is small; when the $\sigma$
components are close to the central $\pi$ component, the line core is
deeper and $W_\lambda$ is quite large. We then experimented with
averages of spectra of similar $W_\lambda$ values and apparently
similar form to obtain averaged spectra of small, intermediate, and
large values of $W_\lambda$ of higher S/N than the individual
spectra. With these averaged spectra of improved S/N we were able to
estimate more reliably the $\sigma - \sigma$ separation, and to then
estimate \bs. We then associated the deduced value of \bs\ for each
averaged spectrum with the average value of $W_\lambda$.

In this way we found that an average of the three spectra with
$W_\lambda \approx 20$\,m\AA\ appear to have $\bs \approx 95$\,kG; an
average of three spectra with $W_\lambda \approx 175$\,m\AA\ have $\bs
\approx 72.5$\,kG; and we estimate that the two averaged spectra with
$W_\lambda \approx 330$\,m\AA\ correspond to $\bs \sim 50$\,kG,
although this value is more uncertain than the other two \bs\
values. If we make the further assumption that $W_\lambda$ is (very
roughly) linearly related to \bs, we can use the simple relation
\begin{equation}
   \bs \approx 50 + 45\frac{330 - W_\lambda}{310}\,{\rm kG}
\end{equation}
to estimate the value of \bs\ to associate with each
spectrum. Estimated values of \bs\ computed in this manner are
reported in Col.~5 of Tab.~\ref{Tab_meas_2359}. We consider that
realistically each of these values is uncertain by roughly 10\,kG.

\section{Search for stellar rotation period}
It is clear from the observational material that both \mwda\ and
\mwdb\ are variable. Judging from a large number of previously studied
MWDs, the variations in the Zeeman line profiles that we observe are
due to the rotation of these stars, and are therefore periodic.

The next step is to search the moment data for each star for periodic
variations. This has been done using the {\sc Fortran} program {\sc
dchi.f}, already employed for many years for period searches on
magnetic field and photometric data for main sequence magnetic stars
\citep[e.g.][]{Neinetal12,Landetal14}. The program starts with a list
of $n$ measurements, for example $\bz_i$, with the associated
uncertainty $\sigma_{i}$ and Modified Julian Date $MJD_i$. For a given
frequency (or period), the
program computes phases of all the data points, and then carries out a
least-squares fit to the observational data using a sine
wave whose zero point, amplitude, and phase are varied. The quality of
the fit is assessed by minimising the value the reduced chi-squared
parameter
\begin{equation}
\chi^2/\nu = \frac{\sum_{i=1}^n [(y_i - y_{\rm fit})/\sigma_{i}]^2}{n-3}
\end{equation}
at that frequency. This quantity is then computed for all independent
frequencies between two given limits (with a sampling spacing chosen
by the user), and the resulting value of $\chi^2/\nu$ of the best fit
found at each period is output as a function of period. If the data
vary periodically in an approximately sinusoidal manner, and the measurement
uncertainties are realistic, the correct period will have a best fit
$\chi^2/\nu$ value of about 1. If the sampling of the data is
sufficiently dense and well spaced, the correct value of the period
may be identified uniquely, but often there are aliasing problems,
especially due to observations spaced at a larger interval of time
than the period.

\subsection{The rotation period of \mwda}
To estimate the period of \mwda\ we searched between 
10 and 0.1\,d for periodic sinusoidal variation of the \bz\
measurements of \mwda\ from 2016 (MJDs starting with 57549), which are
closely spaced. In the range of frequencies searched we found that 
there is only a single frequency that gives a really good sine-wave
fit to the data, at frequency $f = 4.1124 \pm 0.0012$\,cycles/d, or a
period of $P = 0.24317 \pm 0.00007$\,d. Two frequency aliases at around
about 3 and 5\,cy/d, caused by changing the cycle count between
successive nights by $\pm 1$, are strongly rejected because we have
multiple observations of \bz\ on three separate nights.

\begin{figure}[ht!]
  \begin{center}
\scalebox{0.35}{\includegraphics*[trim={0.5cm 1.0cm 2.3cm 2.8cm},clip]{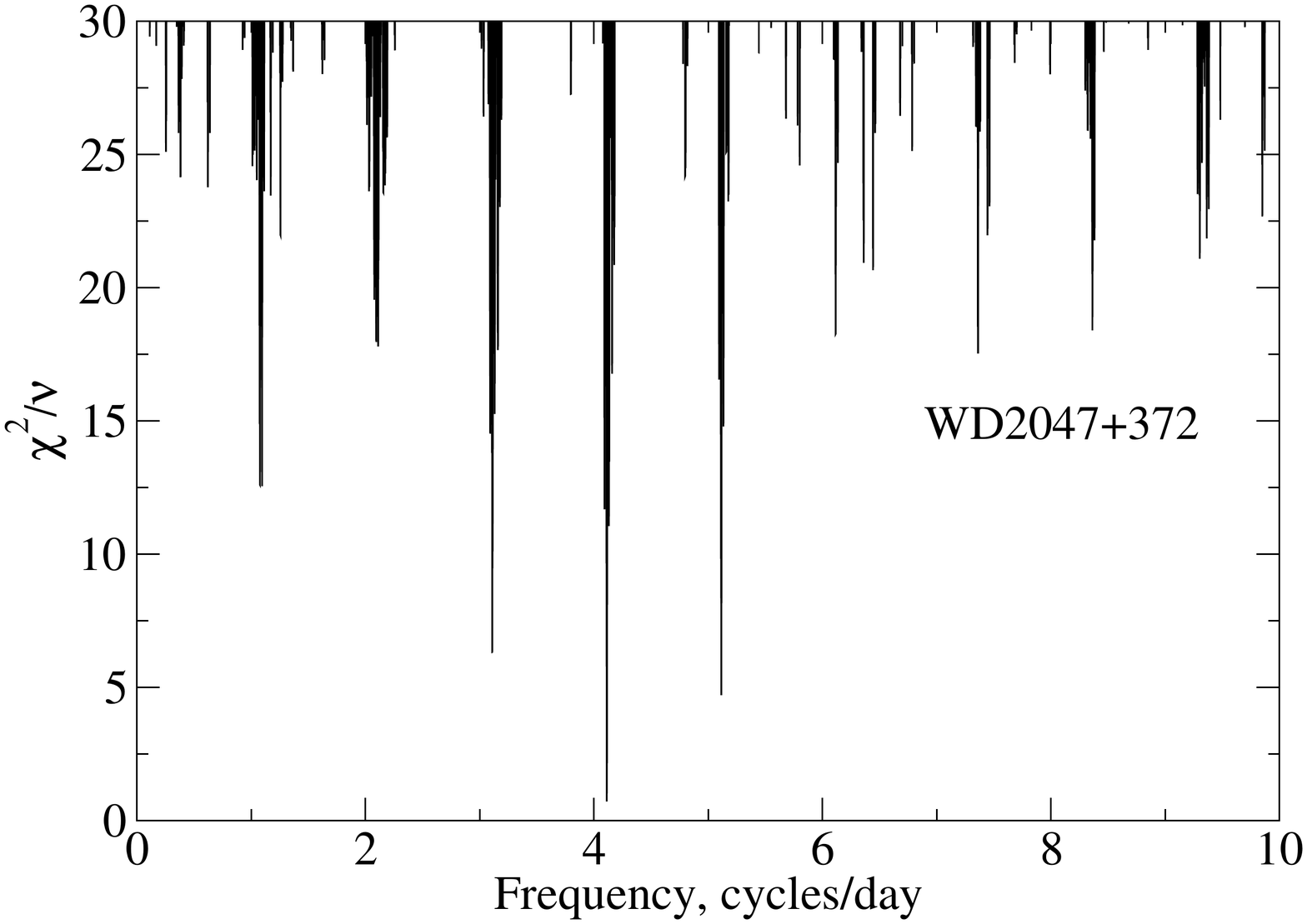}}
\scalebox{0.35}{\includegraphics*[trim={0.1cm 1.0cm 2.9cm 2.8cm},clip]{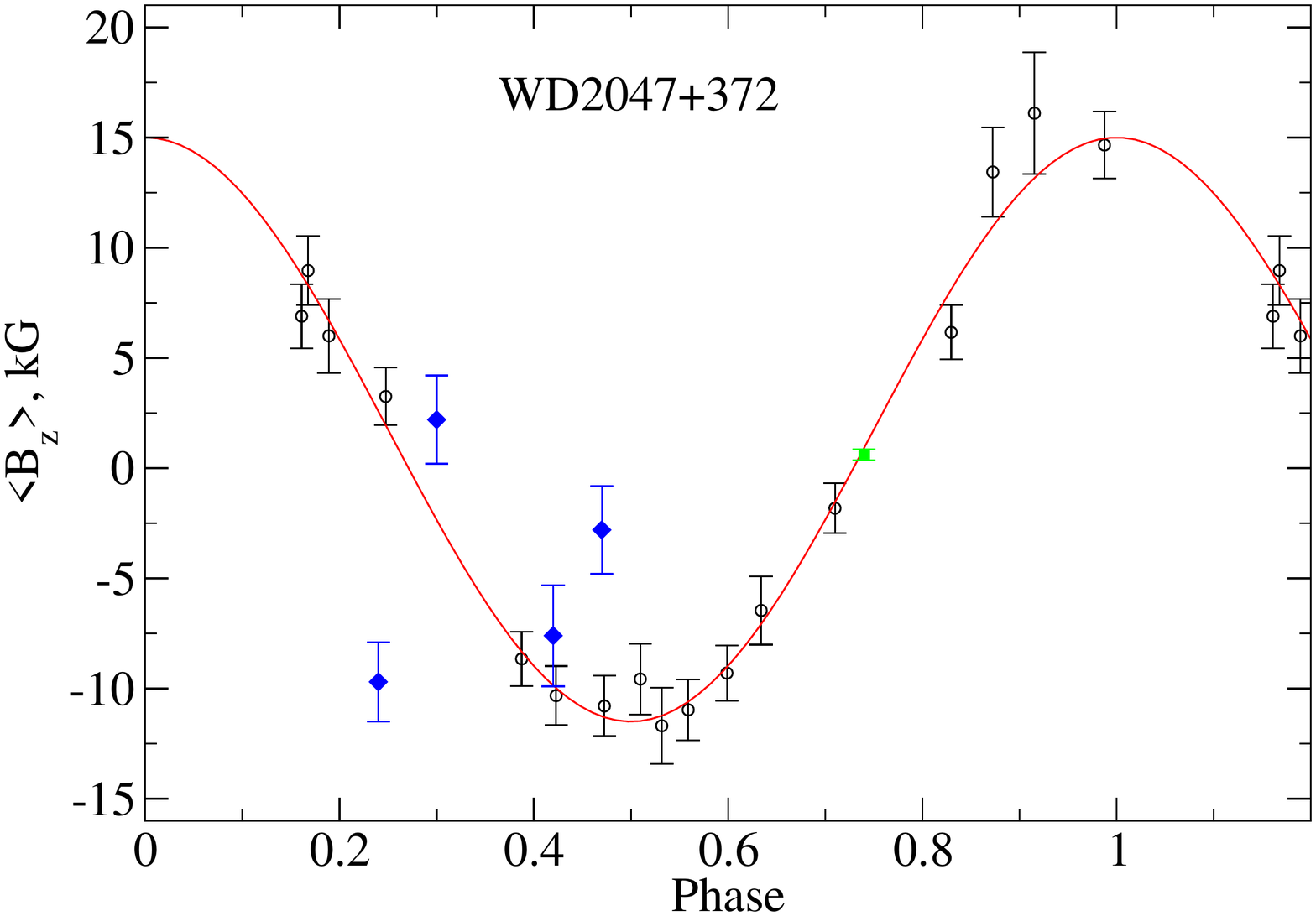}}
\scalebox{0.35}{\includegraphics*[trim={0.1cm 1.0cm 2.3cm 2.8cm},clip]{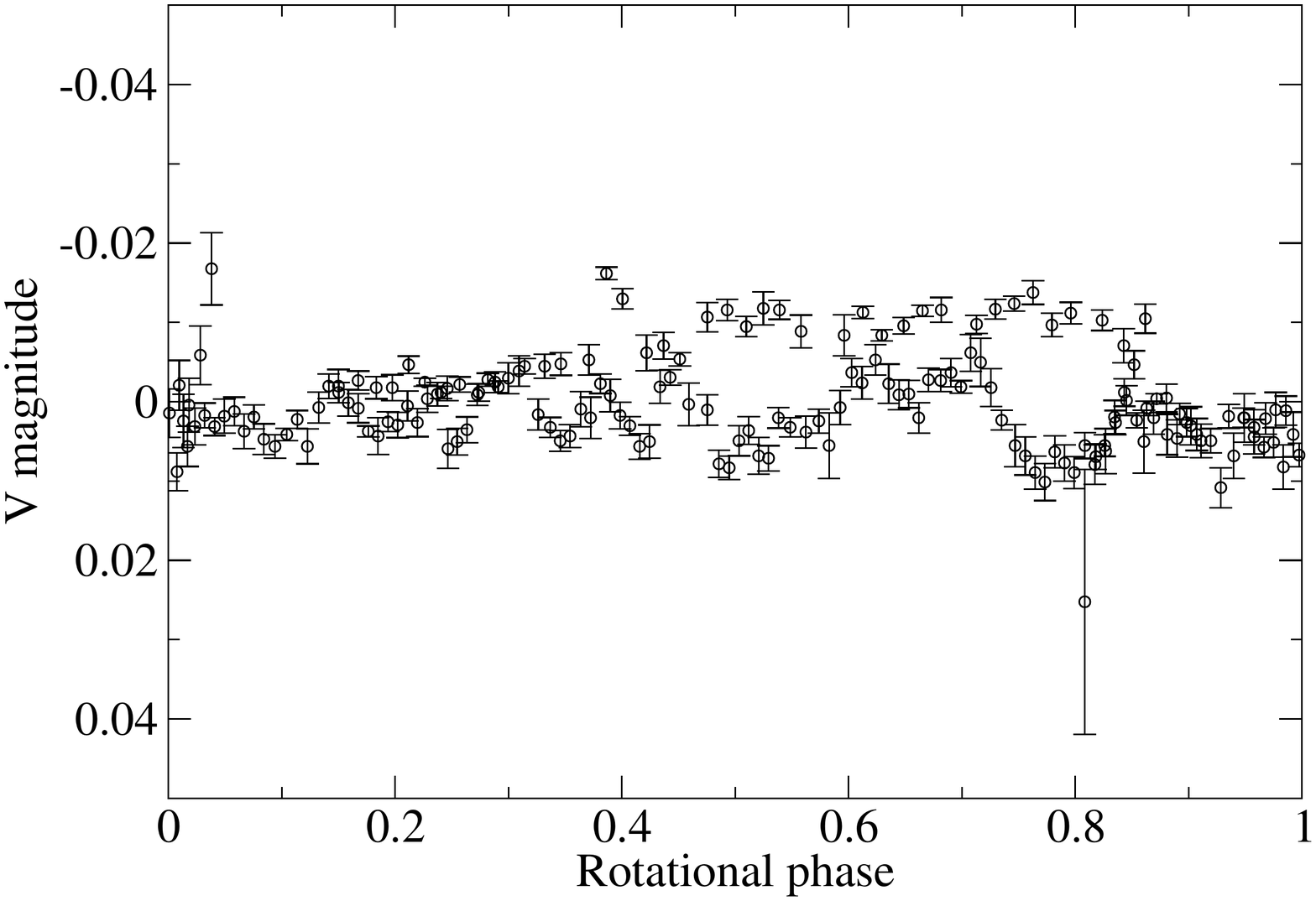}}
\end{center}
\caption{\label{Fig_wd2047_phased} \mwda: {\it Top panel:} chi-squared spectrum
  sine wave fits to all \esp\ \bz\ measurements as a function of
  frequency of field variation. {\it Mid panel:} variation of \bz\ as a function
of phase through Eq.~(\ref{Eq_EPH_2047}), with the origin
of phase at \bz\ maximum (positive) \bz. The very precise measurement at
phase 0.74 (green point) is the first \bz\ value measured, using
ISIS. The four data points with blue filled diamonds are the MSS
data. The smooth curve is the best sine wave fit to the \esp\ \bz\ measurements found from the chi-squared period search. {\it Bottom panel:}
 $m_V$-band flux variations using the same ephemeris as above.
}
\end{figure}
\begin{figure}[ht!]
  \begin{center}
\scalebox{0.35}{\includegraphics*[trim={0.5cm 1.0cm 2.3cm 2.8cm},clip]{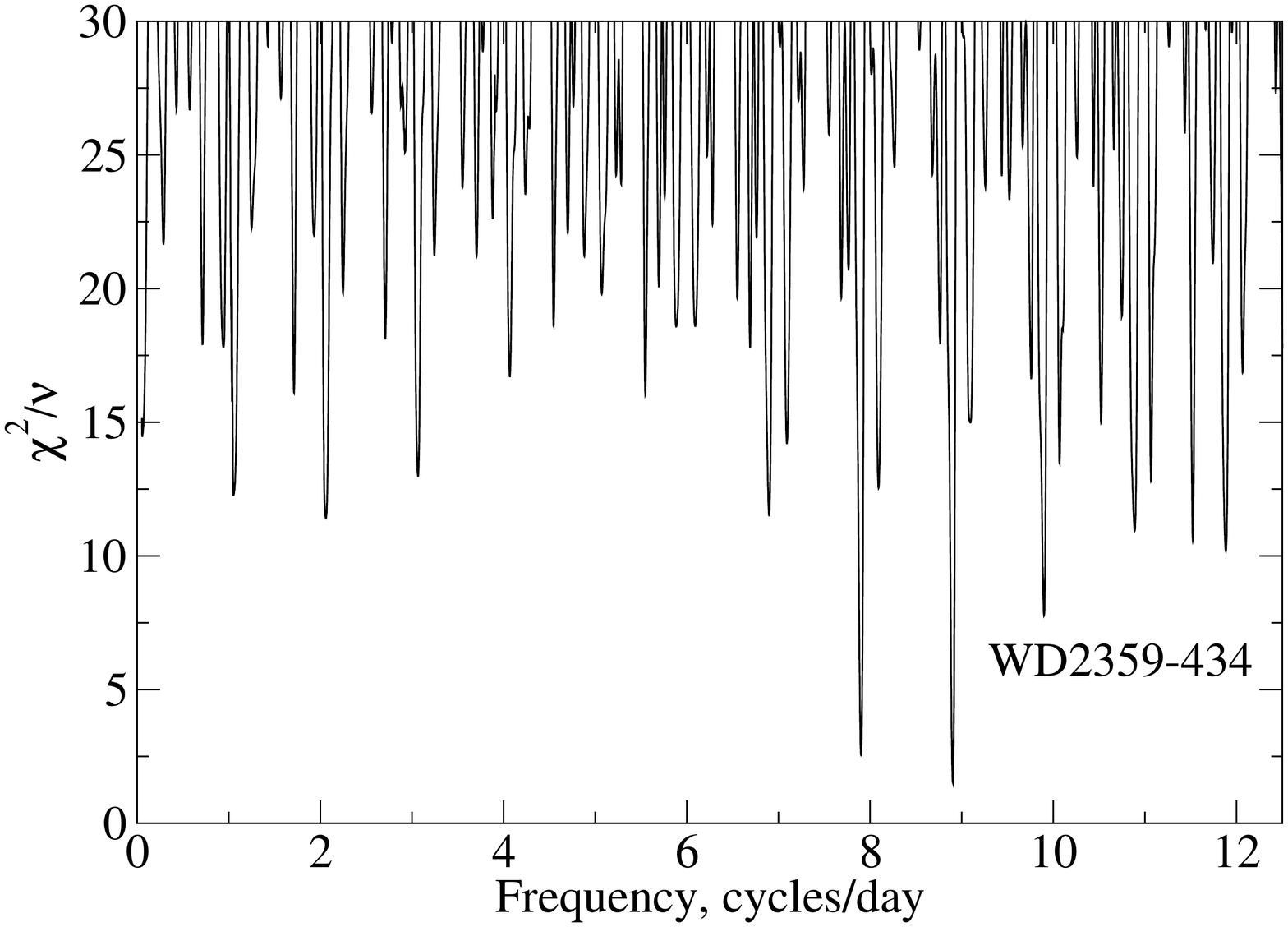}}
\scalebox{0.35}{\includegraphics*[trim={0.2cm 1.0cm 2.3cm 2.8cm},clip]{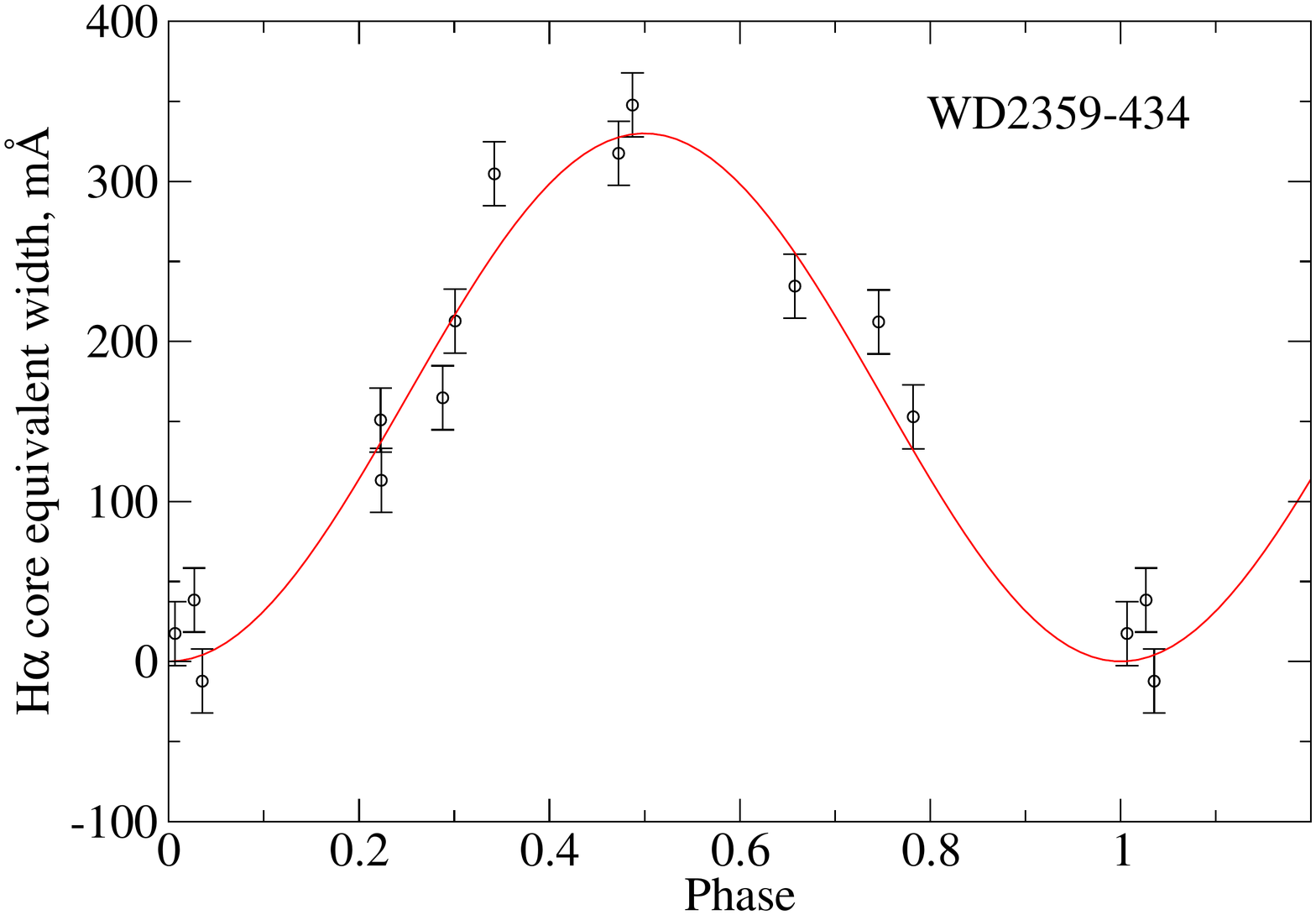}}
\scalebox{0.35}{\includegraphics*[trim={0.5cm 1.0cm 2.3cm 2.8cm},clip]{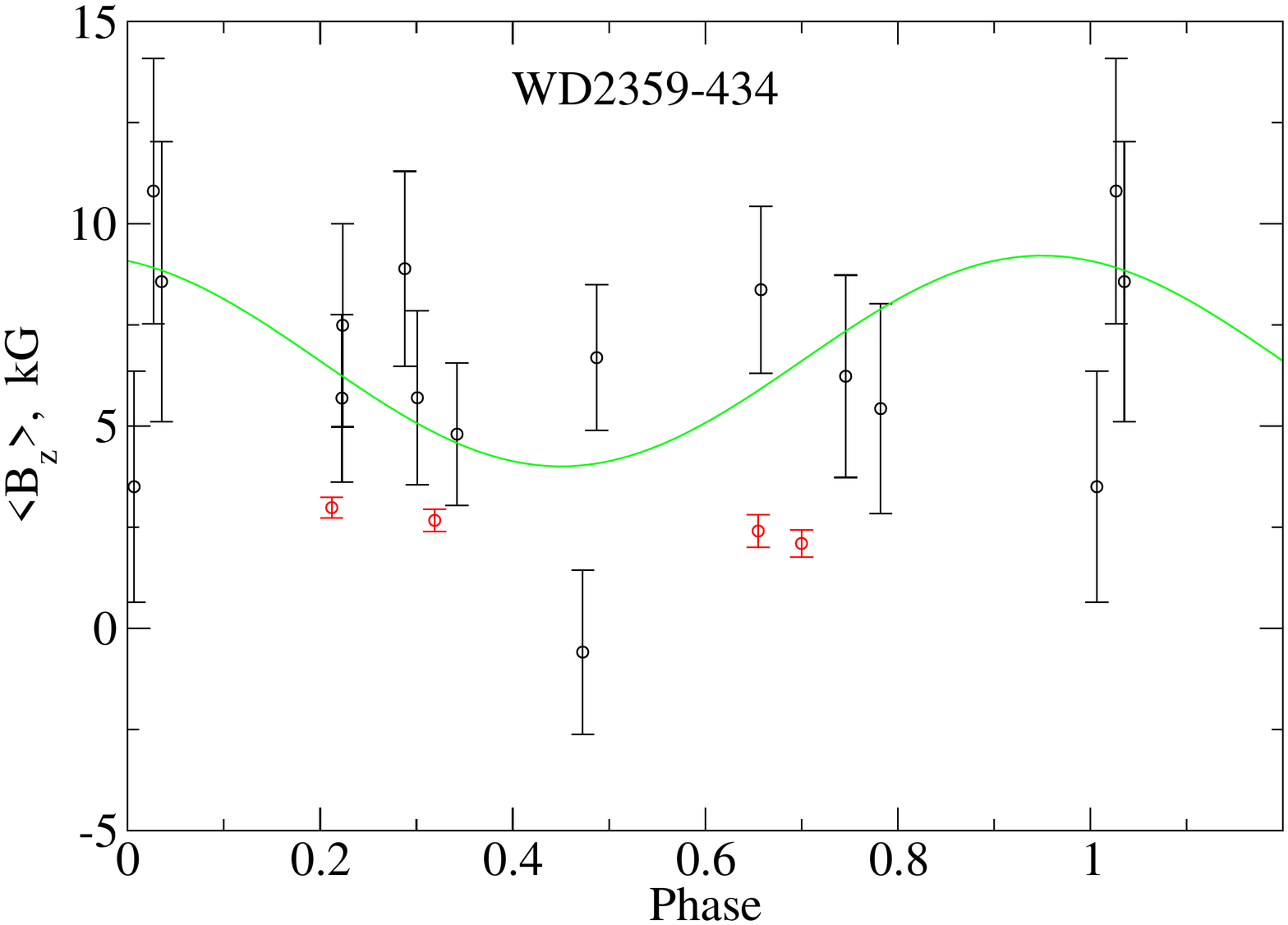}}
\end{center}
\caption{\label{Fig_wd2359_phased} \mwdb. {\it Top panel:} Chi-squared
spectrum of
$W_{\lambda}$ measurements of nine contiguous \esp\ spectra of \mwdb\
taken during ten nights. {\it Mid panel:} Phased variation of \ha\ line
core equivalent width of \mwdb\ using the same ephemeris  of
Eq.~(\ref{Eq_EPH_2359}). The smooth curve is the best fit sine wave.
{\it Bottom panel:}  Variation of \bz\ of
\mwdb\ measured from \esp\ spectra (black symbols) and from FORS spectra
(red symbols), using the same ephemeris as above. The first two FORS1
measurements are not shown because they are too old to be accurately
phased with the epemeris. The smooth curve is the best fit sine wave to the \esp\ data. 
 }
\end{figure}

When the first \esp\ observation, taken about seven months before the
remaining ones, is added to the dataset, the frequency discrimination is
substantially sharper and selects uniquely the frequency $f = 4.11238
\pm 0.00044$, or equivalently the period $P = 0.243168 \pm 0.000024$,
values that are about three times as precise as the values derived
from the shorter data sequence. A portion of the $\chi^2/\nu$ spectrum
of the full data set is shown in the top panel of Fig.~\ref{Fig_wd2047_phased}. Using
the best-fit identified by our program, we find that the rotational
ephemeris of \mwda\ is given by
\begin{equation}
MJD({\rm max}\ \bz) = 57326.209 + (0.243168 \pm 0.000024)*E \;,
\label{Eq_EPH_2047}
\end{equation}
where phase=0\degr\ correponds to the maximum of the mean longitudinal magnetic field.

The \bz\ values for \mwda\ can now be plotted using the best-fit
period. The result is shown in the mid panel of Fig.~\ref{Fig_wd2047_phased}. It is
clear that the variation of \bz\ with phase, using the best period, is
indeed quite sinusoidal. In fact, it seems obvious from the figure, in
which only two data points appear to miss the fitted curve by
substantially more than one standard error, that the uncertainties
computed for the \bz\ values are probably over-estimated by a
significant amount, perhaps by as much as 15--20\%. This is supported
by the minimum value of $\chi^2/\nu = 0.71$, when we expect a value
considerably closer to 1.0.

Because the \bz\ value from the single ISIS measurement was determined
in quite a different manner from those derived from \esp\
measurements, and may measure the field in a somewhat different way
than is done with the \esp\ data, the ISIS \bz\ value was not used in
determining the rotation period of \mwda. However, we have plotted the
ISIS \bz\ value in Fig.~\ref{Fig_wd2047_phased}, where it appears
as a very precise point at about phase 0.74. It will be seen that the
ISIS measurement is in fact in very good agreement with the ensemble
of \esp\ data, and the fact that it did not reveal a value of \bz\
that is significantly different from zero, in spite of very high
precision, is simply due to the moment chosen for observation.

The four \bz\ measurements obtained with MSS have nominal
uncertainties similar to those obtained with \esp. The first and
fourth MSS measurements are in reasonable agreement with \esp\
measurements; the second one has the right sign but is too small in
absolute value.  The third is seriously inconsistent with the variation
of Fig.~\ref{Fig_wd2047_phased} (mid panel), as it deviates from the
curve by about $6\,\sigma$. The discordant results may result from
strongly variable seeing and transparency conditions during the
observations, together with total integrations times that are about
0.15~cycles long, leading to errors in estimated phase of each
measurement of 0.05~cycle or even more. In any case, the four data points, taken
together provide confimation of a field of about $-4.6 \pm
1$\,kG at about phase 0.36, consistent with the \esp\ results. 

The SAO photometric data also were searched for periodicity. Because
the amplitude of variation is comparable to the measurement
uncertainties, a number of acceptable frequencies were found. The
period deduced from \esp\ \bz\ values was one of these, but not the
best. The results of the photometric observations phased with the
ephemeris of Eq.~(\ref{Eq_EPH_2047}) are presented in the lower panel
of Fig.~\ref{Fig_wd2047_phased}, averaged within phase bins of about
0.01 width. The observational uncertainty $\sigma$ is the standard
error of the mean. It is not clear from the figure whether the small
variations (of order 0.01\,mag) are actually coherent and reflect real
stellar photometric variability, or whether they show the difficulties
of trying to detect very small photometric variations under less than
perfect photometric conditions. In any case, these data appear to
extablish that any photometric variability of \mwda\ is at most about
0.02\,mag peak to peak. This is consistent with the very small
magnitude of photometric variations detected for \mwdb, and contrasts
to the considerably larger photmetric variations (up to $\sim
0.1$\,mag) observed in some MWDs with much larger magnetic fields
\citep{Brinetal13}.

\subsection{The rotation period of \mwdb}

While the two archival $I$ spectra of \mwdb\ from the UVES archive
data are almost identical, and suggest little variability, \bz\
measurements from the FORS1 archive data and from our new FORS2 survey
are consistent with some rather modest variations of \mwdb.  The
low-resolution FORS spectra yield quite precise values of \bz, and
indicate that \bz\ is always positive, and varies between about $+2$
and $+4$\,kG (see Tab.~\ref{Tab_meas_2359}). However, these data are
too widely spaced in time to allow us to estimate the stellar rotation
period.  Unfortunately the more closely spaced \esp\ $V/I$ profiles do
not even show strong Zeeman polarisation signatures; while almost all
of the \esp\ \bz\ measurements also report positive fields, these
measurments have uncertainties typically five times larger than the
FORS data, mostly detect a non-zero value of \bz\ only at the
$2-3\sigma$ level, and do not reveal convincing evidence of
variations. These data also are not at all well-suited to period
finding.

In contrast, the \esp\ intensity spectra do show clear variability,
with quite a different pattern of variations compared to those of
\mwda, particularly in the shape of the broad $\sigma$ components of
the Zeeman pattern, and the overall depth of the line core. The
obvious variability of the core of \ha\ provides the most promissing
data for searching for periodicity.

The equivalent width data for the continuous series of spectra between
MJD~57603 and 57613 were searched for good sine wave fits for periods
between 10 and 0.08\,d using {\sc dchi.f}. The best fit was found for
the period $P = 0.11232 \pm 0.00009$\,d, equivalent to a rotation
frequency of $f = 8.9031 \pm 0.0071$\,cycles/d. The time series could
have had significant aliasing problems because all observations had to
be taken (due to the large southern declination) within a period of
about 3~hours each night, but the spread in the local sidereal time of
observation from night to night was large enough to clearly identify
a unique best period. The aliases with frequency differing from the best
frequency by $\pm 1$\,cycle/day are clearly substantially worse. Part
of the $\chi^2/\nu$ spectrum for the nine contiguous spectra is shown
in Fig.~\ref{Fig_wd2359_phased}

After identifying this period we found the results of
\citet{Garyetal13}, reporting the work of an amateur photometry group
which discovered and studied photometric variations of the light from
\mwdb, as described in Sec.~3.5. They have carried out a time series
analysis of their data, and find low-amplitude sinusoidal variation of
the light. Based on observations in 2011 and 2012, \citet{Garyetal13}
propose a period of $P = 2.695022 \pm 0.000014$\,hr $= 0.11229258 \pm
0.00000058$\,d, or $f = 8.905308 \pm 0.000046$\,cycle/d.  Because
they have obtained a large number of light curves, many with several
full cycles per night, this period is unique. Since the publication of
their paper, Gary et al.  have obtained further light curves in 2013,
2014, and 2016, and have confirmed and refined somewhat their period,
as described on their web site\footnote{\tt
http://brucegary.net/WDE/WD2359-434/WD2359-434.htm}. Their period is
fully consistent with, but considerably more accurate than, the period
derived from the 2016 spectropolarimetry of \mwdb.

We then returned to the equivalent width data set, and included the
four equivalent width measurements from 2015 in our period search,
together with the nine from 2016. Because our data do not uniquely
specify the cycle count between October-November 2015 and August 2016,
the allowed frequencies form a comb of acceptable values, of which the
best fits are 8.8946, 8.8981, 8.9017, 8.9052, and
8.9087\,cycles/d. Each of these frequencies is uncertain by about $\pm
0.0002$\,cycles/d. The frequency in best agreement with the results
from the photometry of Gary's group is $f = 8.9052 \pm
0.0002$\,cycles/d ($P = 0.1122939 \pm 0.0000025$\,d). The two periods
are identical within their uncertainties, and it is quite clear that
both variable quantities identify the same rotation period. We adopt
the period emerging from the consistent results of photometry and
spectropolarimetry:
\begin{equation}
  MJD ({\rm min}\ W_\lambda) = 57324.321 + (0.1122926 \pm
0.0000025)*E,
\label{Eq_EPH_2359}
\end{equation}
where the zero point is chosen as the minimum of the equivalent width
(which approximately coincides with the maximum value of \bs).

The variation of the \ha\ equivalent width with this period is shown in
Fig.~\ref{Fig_wd2359_phased}. Within the uncertainties the
variation appears close to sinusoidal.

It is clear from Fig.~\ref{Fig_esp} that, in most
\esp\ spectra, a non-zero signal in the $V$ spectrum is barely
detectable, and this is reflected in the large relative uncertainties
in Tab.~\ref{Tab_meas_2359}. 
However, the magnitude of \bz\ as measured from
\esp\ data seems to be about twice as large as the values from
FORS. This kind of discrepancy, between values of the mean
longitudinal field as measured for a single star by two different
techniques, is also found among upper main sequence magnetic stars
\citep{BorrLand80,Math91}. In this case, the difference in the scale
of \bz\ measurements may arise because a method that uses the core of
\ha\ and a method that uses the wings of higher members of the Balmer
series will certainly have different limb darkening and line weakening
behaviour as a function of colatitude from the centre of the stellar
disk, so that in the two kinds of measurement, various parts of the visible
disk are given different weights in the computation of the observed
values of \bz.


\section{Modelling of the surface magnetic field structure}
\begin{figure*}[ht]
\scalebox{0.47}{\includegraphics*[trim={0.8cm 5.0cm 1.3cm 2.8cm},clip]{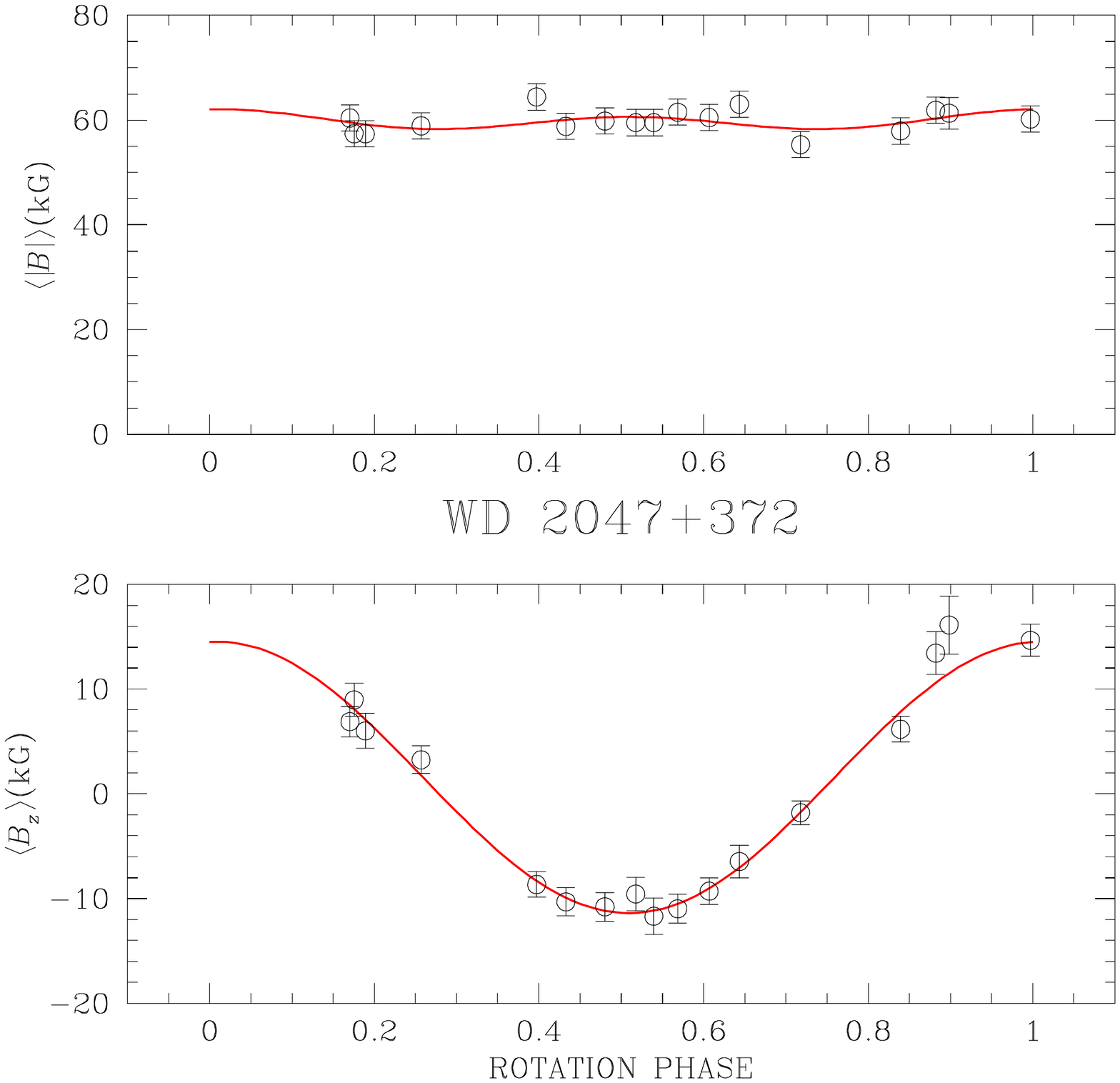}}
\scalebox{0.47}{\includegraphics*[trim={0.8cm 5.0cm 1.3cm 2.8cm},clip]{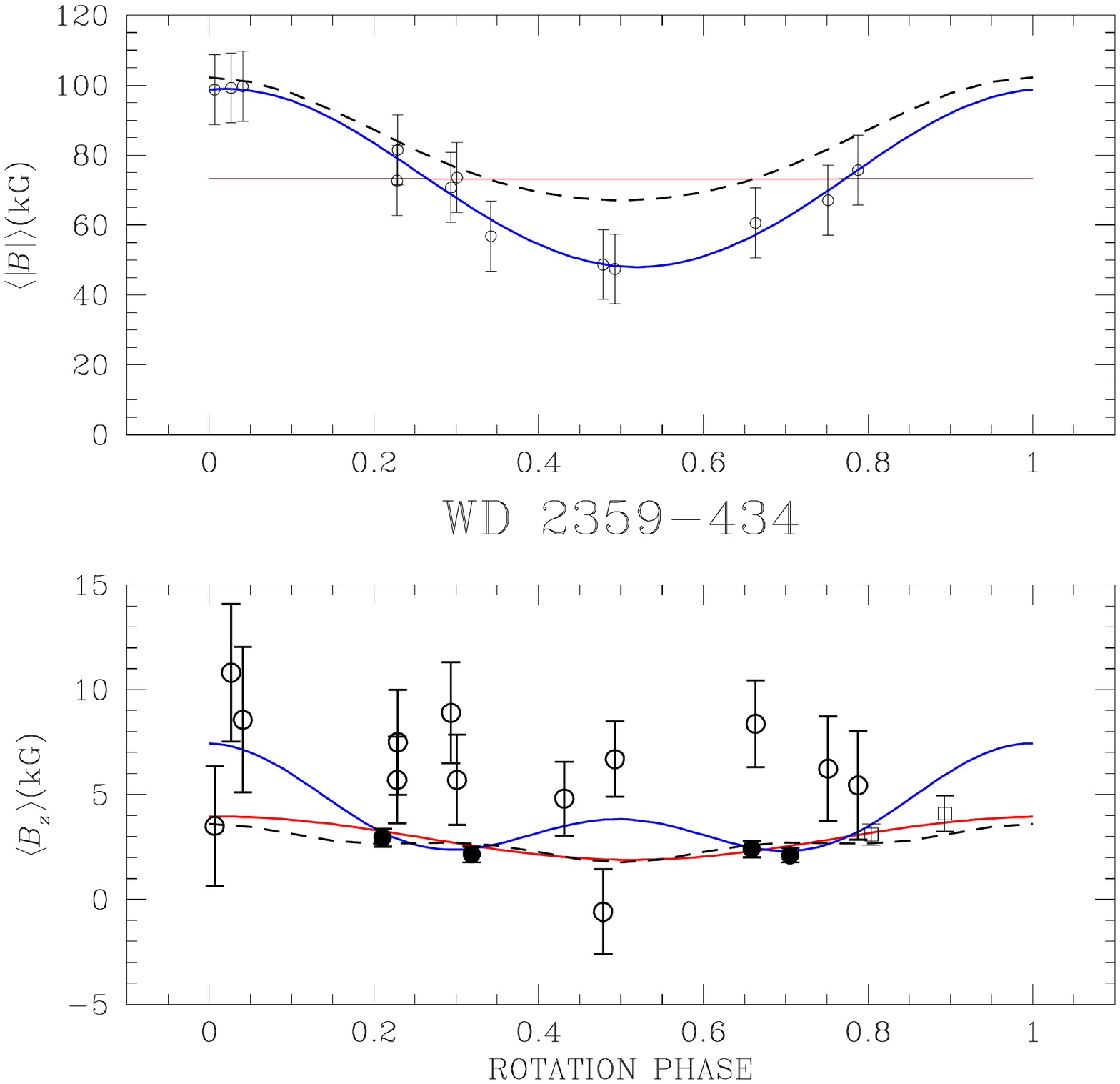}}
\caption{\label{Fig_Models} {\it Left panel:} \mwda: Observed 
  variation of \bz\ and \bs\ fit by a simple dipolar model. {\it Right
  panel:} \mwdb: Observed variation of \bz\ and \bs\ fit by a simple
  dipolar model (thin red lines), a co-linear
  dipole-quadrupole-octupole model (dashed black lines). and by a
  model including the superposition of a dipole with a non linear
  quadrupole (thick solid blue lines), as described in the text.  Both
  FORS data (solid circles) and ESPaDOnS data (empty circles) have
  been used in the fits, with appropriate weights. The two oldest
  FORS1 data points are shown with nominal (but very uncertain) phases
  with open squares, but were not used in finding any of the models. }
\end{figure*}

In this Section we present simple magnetic models that reproduce the
field moment measurements (the mean field modulus and mean
longitudinal magnetic field) of \mwda\ and \mwdb. We experimented with
three surface field multipolar expansions of increasing complexity,
(1) a pure dipolar field, (2) a multipolar expansion with three
axisymmetric, colinear components: dipole, quadrupole, and octupole,
and (3) a multipolar expansion with a dipole and an independently
orientated non-linear quadrupole. For each of these three models it is
straight-forward to compute the distribution over the WD surface of
the magnetic field vector, and from the distribution to predict the
observed field moments \bs\ and \bz.

The simple tilted (oblique) dipolar field has been the simplest model
adopted in many earlier studies of Ap/Bp stars, starting with
\citet{Stibbs50}. It is essentially characterised by a dipolar field
strength at the pole $B_{\rm d}$ and an angle $\beta$ between the
rotation axis and the dipolar axis. The tilt angle between the stellar
rotation axis and the line of sight will be denoted by $i$.

The axisymmetric combination of a colinear dipole, linear quadrupole
and linear octupole was already used by the fitting programs described
by
\citet{LandMath00}. In addition to the angles $i$ and $\beta$ and to
the dipolar field strength $B_{\rm d}$, the additional model
parameters are the quadrupole and octupole strength $B_{\rm q}$, and
$B_{\rm o}$, respectively.  This parametrisation is rather similar to
the one used by \citet{Kueletal09} to model single snapshot intensity
spectra of magnetic white dwarfs detected in the SDSS data set. Their
basic model, a dipole either centred in the star (as all our models
are) or decentred along the dipole axis, is quite similar in global
surface field structure to the sum of an aligned linear dipole and
quadrupole, with their decentring parameter playing roughly the same
role as the ratio of $B_{\rm q}/B_{\rm d}$ in the multipole expansion
\citep[as analytically demonstrated by][]{Deretal79}. Such models can
roughly describe topologically dipolar fields, and can also describe a
large area of roughly constant local field $\vert B \vert$ around one
pole, with a flux concentration (a ``magnetic spot'') at the other
pole. The addition of a colinear octupole basically allows us to
modify (e.g. reduce) the large pole-to-equator field strength
difference of the pure dipole without introducing a spot, which may be
required for mapping based on both $I$ and $V$ spectra phased around a
full WD rotation cycle, rather than $I$ only from a single snapshot.

The non axisymmetric model includes the contribution of a dipole and a
non-linear quadrupole, as described by \citet{Bagetal96} and
\citet{Lanetal98}, and the fitting  algorithm is fully described in 
\citet{Bagetal00}.  This model has three more free parameters than the
previous model, because even though it does not include an octupolar
component, the orientation of the quadrupole is defined by four
independent angles, ($\beta_1,\gamma_1$) and ($\beta_2,\gamma_2$).

\subsection{\mwda}\label{Sect_MWDA}

Finding a model of \mwda\ consistent with our measurements is
relatively straightforward. A simple dipolar model with $i = 27.0\degr
\pm 1.5\degr$, $\beta=86.5\degr \pm 1\degr$ and $B_{\rm d} = 91.8 \pm
0.8$\,kG reproduces the \bz\ and \bs\ observations with a reduced
$\chi^2 \sim 0.8$ (see left panels of Fig.~\ref{Fig_Models}). The
$\chi^2$ value of slightly less than 1.0 suggests that our error bars are slightly
overestimated. Within the framework of a simple dipolar model, the
solution is unique, although it is not possible to distinguish the
degenerate solutions in which the ($i,\beta$) values are exchanged as
follows: ($i,\beta$), ($\beta,i$),($180\degr-i,180\degr-\beta$),
($180\degr-\beta,180\degr-i$) \citep[see][]{Lanetal98}. The parameter
values are mildly sensitive to the assumed behaviour of the model
stars, particularly the assumed limb darkening and line weakening
towards the limb, but changing the assumptions about these effects has
only a small effect on the resulting model.

It is found that an axisymmetric model with a mix of dipole and
(opposing) octupole fits the moment data as well as the simple dipole. A
successful model of this type has $B_{\rm d} = 80$\,kG, $B_{\rm o} =
-40$\,kG, and ($i,\beta$) $\simeq$ (85\degr,33\degr). In comparison with
the simple dipole model, which has a polar field strength of 92\,kG,
declining to about 46\,kG at the equator, the dipole-octupole field
has a polar field of 40\,kG, rising to about 80\,kG around the
equator. This difference makes it clear that the moment data do not
strongly constrain possible modest variations of local field strength
from pole to equator to opposite pole. Comparison of the two field
models also shows that using the (net) polar field strength as a
measure of the typical MWD field is not nearly as robust as using
actual values of \bs\ if these are measureable.

\subsection{\mwdb}

\begin{figure*}[ht]
\scalebox{0.95}{\includegraphics {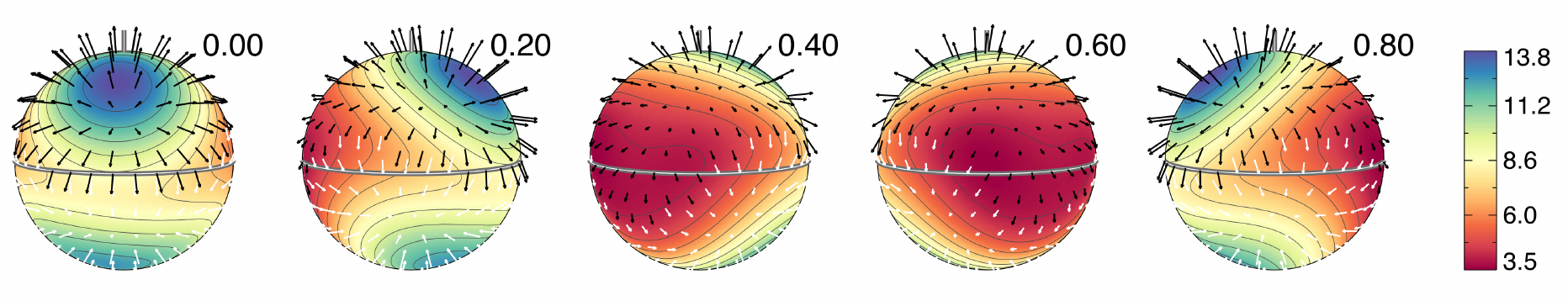}}
\caption{{\label{Fig_wd2359_field_model}} The distribution of magnetic field
over the surface the dipole-non-aligned quadrupole model of \mwdb, as
seen at five successive phases (left to right: phases 0.0, 0.2, 0.4,
0.6 and 0.8). Black arrows represent outward field, white arrows
inward field. The axis of rotation is a small white line segment close
to the top of each sphere. The scale at right is in units of 10\, kG (e.g. 13.8 = 138\,kG).}
\end{figure*}


For \mwdb\ the situation is more complicated. The longitudinal field
\bz\ as measured using FORS appears to be systematically smaller (by
about a factor of two) compared to the values measured with
\esp. Furthermore, although it is clear that \bs\ is variable
with rotation phase, and that its order of magnitude of \bs\ is $\sim
50 - 100$\,kG, its estimate is not very accurate. A third peculiarity
is the small ratio of $\bz/\bs$, which is clearly below about 0.1 at
all phases in the FORS set of our \bz\ measurements. Within the
framework of a dipolar model (or any axysimmetric model) this feature
can be described by a magnetic field characterised by a strong field
at the pole, with the magnetic axis always nearly perpendicular to the
line of sight as the star rotates, so that the longitudinal field,
averaged over the stellar disk, is very low compared to the mean field
modulus. This could be obtained either by ($i,\beta$) $\simeq$
(0\degr,90\degr) or ($i,\beta$) $\simeq$ (90\degr,0\degr).  Because
the star shows rather strong spectroscopic variability, it is unlikely
that the rotation axis is seen pole-on, therefore our favourite
solution is ($i,\beta$) $\simeq$ (90\degr,0\degr).  The best-fit
obtained with a pure dipolar model is shown with red lines in the
right panels of Fig.~\ref{Fig_Models}, and is obtained with
($i,\beta$) $\sim (87\degr,1.5\degr)$, and a dipolar strength $B_{\rm
p} \sim 145$\,kG. Clearly this model cannot account for the large
changes in \bs\ that are observed as the star rotates. 

A co-linear dipole-quadrupole-octupole model can be found that more
nearly reproduces the observed weakness of \bz\ and the large and
variable \bs\ field, by making use of the fact that both quadrupole
and octupole also produce weakly non-zero \bz\ values. It is found
that a model with $i = 60^\circ$, $\beta = 40^\circ$, $B_{\rm d} =
+20$\,kG and $B_{\rm q} = B_{\rm o} = -100$\,kG has a closer
resemblance to the observed field moment variations than a simple
dipole. The field moments predicted by this model are shown in the
right panels of Fig.~\ref{Fig_Models} as black dashed lines.

A much better model that also accounts the observed variability of the
mean field modulus is obtained by considering the superposition of a
dipole and a non-axisymmetric quadrupole, and is shown in the right
panels of Fig.~\ref{Fig_Models} with blue thick solid lines.  This
model has the rotation axis inclined to the line of sight by $i =
83^\circ$, a dipolar field ($\sim 107$\,kG) tilted at 2\degr\ with
respect to the rotation axis, and a 63\,kG, quadrupolar component
whose orientation is defined by the angles ($\beta_1,\gamma_1$) =
(43\degr, 14\degr), and ($\beta_2,\gamma_2$) = (71\degr,
342\degr). However, the non-axisymmetric model is not unique, so this
model should be considered as a reasonable example of a simple
multipolar field that reproduces the field moment data. The
distribution of $|B|$ over the surface of the model is illustrated in
Fig.~\ref{Fig_wd2359_field_model}.

\section{Spectral synthesis}

Clearly our simple models of \mwda\ and \mwdb, found by fitting only
\bz\ and \bs\ variations, are expected to be at best rather simplified
approximations to the large-scale structures of the real surface
field distributions. They are also not unique except within strictly 
limited families of models. However, in principle we have further
information about the field surface geometry in the actual observed
$I$ and $V$ line profiles from which the field moments are
derived. These original profiles could be compared with line profiles
calculated on the basis of the model. Such comparisons could provide
useful tests of the whether our models are usefully realistic.
Furthermore, discrepancies between observed and computed $I$ and $V$
line profiles can be used to improve the model surface field
distributions.

With a code for spectral synthesis of the magnetic atmospheres of
degenerate stars, one could think of performing modelling by inverting
the observed Stokes profiles, as is done for instance to map the
magnetic field geometries of a number of main sequence magnetic Ap/Bp
stars \citep[e.g.][]{Kochetal11,Silvetal15}, and also to model
global field geometry of some strong-field WDs
\citep[e.g.][]{Jord92,Euchetal05,Kueletal09}. However, such a project
encounters a number of specific problems. MWDs are much fainter than the
main sequence stars so far studied, so the time resolution and \snr\
of the data are generally lower. Only a few spectral lines are present
in the entire optical spectrum. For most known MWDs no polarimetric
data at all are available. Finally, mapping of main sequence magnetic
stars is greatly aided by the fact that most show significant
rotational line broadening, allowing different parts of the observed
line profile to be associated with different locations on the
projected visible disk of the star. This is generally not possible for
MWDs, whose line cores are not significantly broadened by rotation,
but are broadened by pressure (to a full width of more than 1\,\AA).

Furthermore, computing a model of the observed line profiles faces a number of
important obstacles. A major set of problems concerns the \ha\ line
cores on which we have relied for much of the data discussed in this
paper. This line core is strongly affected by departure from LTE, and
is not modelled adequately by LTE codes: computed LTE cores are much
shallower than observed \citep{GreePete73,KoesHerr88}. Furthermore,
the correct computation of even the inner line wings requires
specification of both absorptive and dichroic line coefficients
\citep{Beck69,Witt74}. Furthermore, even for the relatively weak fields of
the MWDs of interest here, the actual form of the Stark broadening of
the H lines in the presence of a magnetic field is unknown; we must
use rough approximations \citep{Jord92,Frieetal94}. Nevertheless, we
have carried out some experimental computations of the line cores of
\ha\ and H$\beta$ using the code {\sc zeeman.f}
\citep{Land88} to see whether these can help us to test and improve
the models.

\begin{figure}[ht]
\scalebox{0.36}{\includegraphics{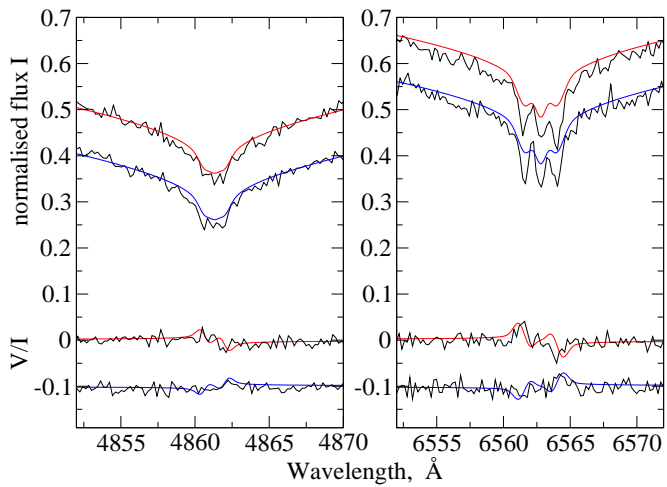}}
\caption{\label{Fig_wd2047_spec_synth} Line synthesis of H$\beta$
(left panel) and \ha\ (right panel) for \mwda. Observed black $I$ and
$V$ curves paired with red models are correctly normalised, lower
curves paired with blue models have been shifted downward by 0.1 for
clarity. Upper and lower spectra from phases 0.98 (\bz\ maximum) and
0.53 (\bz\ minimum) respectively. }
\end{figure}

For synthesis of \ha\ and H$\beta$ in \mwda, we have simply adjusted
the Lorentz damping constants of the line in such a way as to fit
approximately the wings close to the line core. Synthesis of \ha\
using the simple dipolar model of Sect.~\ref{Sect_MWDA} produces a line core
that is considerably shallower than the observed core, but has the
right qualitative width and shape. For H$\beta$, which does not have a
deep non-LTE core, the computed profile provides an acceptable fit to
both the depth and shape of the observed profile. The computed
amplitude and shape of the $V$ signal is also compatible with the (low
\snr) observed $V$ signature. Comparison of $I$ and $V$ data with
synthetic spectra for two representative phases are shown in
Fig.~\ref{Fig_wd2047_spec_synth}. Using the dipole-octupole model
mentioned in the previous Sec. changes the computed line profiles only
marginally, as the changes in the local field strengths and
displacements of the $\sigma$ components is small compared to the
pressure-broadened line core width. Thus all the data suggest
that the magnetic field of \mwda\ may well be axisymmetric, with a
local field strength that does not vary around a mean of about 60\,kG
by more than perhaps $\pm 30$ or 40\,\%, but we have no constraints on
local field variations within these limits. 

Synthesis of Balmer lines in WDs of $\te \la 10000$\,K with our code
is currently very unsatisfactory; the deep pointed cores of \ha\ and
H$\beta$ in non-magnetic WDs are not reproduced even approximately by
our computations. In spite of this, synthesis does generate the
relatively shallow line cores of \mwdb\ fairly successfully. However,
in order to test the models against the subtle shape changes found in the
\ha\ $I$ spectra of \mwdb, we need to be able to compute more
believable \ha\ line profiles than we are able to at present. This is
unfortunate, as the varying shape of the \ha\ line cores clearly
provides more information about the field than is captured by the
field moments. In particular, the great breadth of the Zeeman $\sigma$
components in \mwdb\ tells us that the dispersion of the local field
strength over the visible hemisphere is very much larger, especially
around phases of \bs\ maximum, than is the case for \mwda. This dispersion suggests that there may also be a substantial amount of $V$ signal cancellation over the observed hemisphere, leading to the different values of \bz\ deduced from FORS and \esp\ observations. 

\section{Discussion and Conclusions}

We have reported detection of obvious magnetic variability in the
the cores of \ha\ of two weak field magnetic white dwarfs:
\mwda\ and \mwdb.  For \mwda\ the shape of the Zeeman split line core
is almost constant, but polarimetry reveals that the line of sight
component of the field, \bz, varies strongly. In contrast, for
\mwdb\ the polarimetric signal is weak and varies rather little, but
the line profile in intensity reveals large changes in the shape of
the Zeeman $\sigma$ components, reflecting strong changes in the field
strength distribution on opposite sides of this star.

For each of the stars, the measurements reveal a unique stellar rotation
period.

The rotation period of \mwda, 0.24317\,d, is detected for the first
time in our data. Furthermore, the internal consistency of the
\bz\ measurements used to determine the period strongly confirms the
stability and precision of the \esp\ spectra (and even suggests that
our assigned standard errors of measurement are somewhat too
conservative).  For this star we report new photometry, which does not
show clear periodic variation with the rotation period, but which does
limit any variations to a semi-amplitude of about 0.01~mag.  

The newly discovered magnetic variability of \mwdb\ leads to a
rotation period of 0.112292\,d, a period found independently from our
spectra and from photometry, from which it is found that
recent light minima of
\mwdb\ occurred on MJD\,57631.785 (phase 0.09 of our ephemeris), and
on MJD\,57635.833 (phase 0.14). (Probably the small phase difference
arises from the uncertainties of fitting sine waves to data strings of
individual nights.) It thus appears that minimum light occurs near or
shortly after the closest approach of the line of sight to the region
of largest field strength, while maximum light occurs about the time
when the line of sight intersects the region of relatively weak field.

We have used the new data to obtain simple models of the magnetic
field configurations of both the MWDs studied here.

The variations with rotational phase of \bz\ and \bs\ observed for
\mwda\ suggest a field structure that is globally dipolar in
structure. It appears that the field axis is inclined to the rotation
axis by $\sim 30\degr$, and that the rotation axis is inclined almost
normal to the line of sight, at about $i = 87\degr$ (or,
alternatively, $i \sim 30\degr$ and $\beta = 87\degr$)).  The local
field strength $|B|$ appears to be roughly constant over the surface at
about 60\,kG. This model appears to be consistent with all available
field moment data, and leads to synthesized \ha\ and H$\beta$ $I$ and
$V$ line profiles that closely resemble to observed ones. We are led
to conclude that this model provides a plausible low-resolution
map of the surface field of \mwda, the first such map derived for a
super-weak field MWD. \citep[This field structure, with rather uniform
\bs, is remiscent of the very uniform value deduced for the 200\,MG
field of HE\,0241-0155 by][]{Reimetal04}.


\begin{table*}[th]
\begin{center}
  \caption{Confirmed MWDs with magnetic fields of less than about
  200\,kG }
\label{Tab_wkfld_mwds}
\begin{tabular}{ccccccccc  }\hline 
WD designation & Name & $m_V$    & Spectral & \te   & Typical \bs & Period & Photometry\\
           &            &  (mag) & class    & (K)   & kG      & (days) & \\
\hline
WD0446-789 & BPM 3523   & 13.47  & DA2.1    & 24440 & 15      & V      & ND \\
WD2105-820 & LTT 8381   & 13.59  & DA4.8    & 10600 & 43      & NV     & ND \\
WD2047+372 & LTT 16093  & 13.40  & DA3.4    & 14710 & 60      & 0.2432 & V? \\
WD1653+385 & NLTT 43806 & 15.86  & DZA      & 5833  & 70      & ND     & ND \\
WD0257+080 & LHS 5064   & 15.90  & DA7.8    & 6340  & 90      & ND     & ND \\
WD2359-434 & LTT 9857   & 13.05  & DAP5.8   & 8650  & 50--150 & 0.1123 & V\\
WD0322-019 & G77-50     & 16.22  & DAZ      & 5195  & 120     & 29.8 or 1.03 &
ND \\
\hline
\end{tabular}
\tablefoot{$m_V$ and spectral class from Simbad database; \te\ from \citet{Gianetal11} and \citet{Giametal12}; typical \bs\ for WD0446-789, WD2047+372, and WD2359-434 from our own estimates, for WD2105-820 from \citet{Landetal12}, for WD1653+385 from \citet{Zucketal11}, for WD0257+080 from \citet{Koesetal09}, and for WD0322-019 from \citet{Farietal11}; for period from this work and \citet{Farietal11}. In the period and photometry columns, ND: no data; NV: not variable; V: variable}
\end{center}
\end{table*}

The \ha\ line core of \mwdb\ shows clear Zeeman splitting, with
strongly variable width and depth, indicating the presence of a
strongly variable field of the order of 50--100\,kG. We have found
that the magnetic field of \mwdb\ is very probably not axisymmetric,
and we have identified a simple multipolar model that reproduces the
variations of \bz\ and \bs. Our \ha\ line profiles clearly contain
more information than is captured by the field moments. However,
modelling of the actual \ha\ line cores of this star, to test and
refine possible field models, still presents major challenges.

Nevertheless, it is clear that the magnetic field of \mwdb\ does {\em not}
resemble that of a dipole centred in the star, and in fact does not
even have an axis of symmetry. The field shows very large variation
over the surface (a factor of order three or more), especially when the
stronger field region is visible. This field structure, although not
uniquely modelled, is clearly very different from that of
\mwda, on which the local field does not appear to vary much in strength over
the visible surface. 

The origin of the major differences between these two surface field
geometries is not clear at present. In fact, starting from the view
that the fields of these two stars are fossil fields, one would expect
that the complexity of the global field structure would decrease with
increasing age, as the higher order multipole components decay
ohmically more rapidly than lower order terms
\citep[Sec.~1.3]{Cowl76}. In this case, if the two stars started their
lives with fields of comparable complexity, we would expect that
\mwdb, which is four times older than \mwda, would have the simpler
field, perhaps even a roughly dipolar field. Instead, we find the
exact opposite situation: \mwdb\ has a much more complex surface field
than \mwda, whose field shows no indication of significant departure
from a simple dipole.

Note that the field structure modelling carried out here is unusual in
that it is based on spectropolarimetry carried out throughout known
rotation periods. Although there has been a lot of modelling of
spectroscopy and spectropolarimetry of MWDs to obtain information
about the surface magnetic field distributions, this has generally been
based only on one or a few snapshot observations.  Models based on
observations spread through a rotation cycle, which are considerably better
constrainted than those based on snapshots, have been published for 
only a few MWDs, including 
WD0009+501 \citep{Valyetal05,Valyetal06,Valeetal15}, Feige 7
\citep{Achietal92}, RE\,J0317-853 \citep{Burletal99}, PG\,1015+014 
\citep{Euchetal06}, PG\,1031+234 \citep{Schmetal86}, HE\,1045-0908
\citep{Euchetal05}, and WD1953-011 \citep{Maxtetal00,Valyetal08}. Except for WD0009+501, all
of these MWDs have fields of 10\,MG or more. Our models are the first
based on full-phase data for any extremely weak-field MWDs.

It is of interest to consider the two MWDs studied here in the context
of the seven MWDs known to have fields \bs\ below about 200\,kG (the
MWDs closest to the threshold of field detection), which are
summarised in Table~\ref{Tab_wkfld_mwds}. Until recently, little more
was known about the fields of these stars beyond a snapshot value for
each star of \bs\ or \bz. With this paper we have considerably
increased the amount of information about two very different members
of this highly interesting set. It will be of great interest to obtain
the necessary data for modelling a few more of these stars.

One encouraging aspect of the summary of super-weak field stars is
already clear from Tab.~\ref{Tab_wkfld_mwds}, namely that discovery of
this field strength range among bright WDs must still be very
incomplete, as four (more than half) of the stars in
Tab.~\ref{Tab_wkfld_mwds}) are brighter than $m_V = 13.5$. No stars in
the magnitude range 13.5 to 15.5 are found in the Table, although we
would expect of order 10 times as many super-weak MWDs in this
magnitude interval as are found brighter than 13.5. Many of these
relatively bright but undiscovered super-weak field MWDs must be among
the numerous known WDs in this magnitude range. Thus continued
searches with high-resolution instruments are likely to enlarge the
relatively bright, easily studied weak-field sample substantially.

The photometric variations found in one of the two super-weak MWDs
studied here, and our upper limit for the other, are consistent with
similar low-amplitude variations observed in two other weak field
MWDs, WD\,0009+501 (= G217-37) and WD\,0853+163 (= LB\,8915). However,
the amplitudes of the variations of the two stars studied here, less
than 1\,\% peak-to-peak amplitude, are somewhat smaller than the
variations observed in the two other stars, and considerably smaller
than those reported for a few MWDs with MG fields
\citep{Brinetal13}. The explanation for very low-amplitude light
variability in such weakly magnetic WDs is not obvious. 

The results presented here clearly demonstrate that a 4-m class
telescope equipped with a medium or high resolution, high efficiency
spectrograph (or even better, a spectropolarimeter) can provide data
from the core of \ha\ that reveal aspects of white dwarf magnetism
that are not accesible, especially for the weakest fields, with the
conventional low resolution instruments normally used to observe white
dwarfs. Because it is not clear in advance whether the $I$ or the $V$
spectrum of a weak-field MWD will prove more useful for studying
variability and constraining field models, it is particularly valuable
to use a spectropolarimeter.

\begin{acknowledgements}

  We thank the Queue Service Observing team at the CFHT for
  outstanding help with interactive scheduling of our observations,
  and for providing us with long, consistent data sets. We thank Dr
  Oleg Kochukhov (Uppsala University) for very helpful advice and
  a macro for making Fig.~6.
  JDL acknowledges financial support from the Natural Sciences and
  Engineering Research Council of Canada. 
  GV and AV acknowledges the support from the Russian Foundation for Basic
  Research (RFBR grant N15-02-05183).
  This research has made use of the VizieR catalogue access tool, CDS,
  Strasbourg, France. The original description of the VizieR service
  was published in A\&AS 143, 23.

\end{acknowledgements}

\bibliography{Biblio}

\begin{thebibliography}{62}
\expandafter\ifx\csname natexlab\endcsname\relax\def\natexlab#1{#1}\fi

\bibitem[{{Achilleos} {et~al.}(1992){Achilleos}, {Wickramasinghe}, {Liebert},
  {Saffer}, \& {Grauer}}]{Achietal92}
{Achilleos}, N., {Wickramasinghe}, D.~T., {Liebert}, J., {Saffer}, R.~A., \&
  {Grauer}, A.~D. 1992, \apj, 396, 273

\bibitem[{{Angel} \& {Landstreet}(1971)}]{AngeLand71}
{Angel}, J.~R.~P. \& {Landstreet}, J.~D. 1971, \apjl, 164, L15

\bibitem[{{Appenzeller} {et~al.}(1998){Appenzeller}, {Fricke}, {F{\"u}rtig},
  {G{\"a}ssler}, {H{\"a}fner}, {Harke}, {Hess}, {Hummel}, {J{\"u}rgens},
  {Kudritzki}, {Mantel}, {Meisl}, {Muschielok}, {Nicklas}, {Rupprecht},
  {Seifert}, {Stahl}, {Szeifert}, \& {Tarantik}}]{Appetal98}
{Appenzeller}, I., {Fricke}, K., {F{\"u}rtig}, W., {et~al.} 1998, The
  Messenger, 94, 1

\bibitem[{{Aznar Cuadrado} {et~al.}(2004){Aznar Cuadrado}, {Jordan},
  {Napiwotzki}, {Schmid}, {Solanki}, \& {Mathys}}]{Aznaetal04}
{Aznar Cuadrado}, R., {Jordan}, S., {Napiwotzki}, R., {et~al.} 2004, \aap, 423,
  1081

\bibitem[{{Bagnulo} {et~al.}(2015){Bagnulo}, {Fossati}, {Landstreet}, \&
  {Izzo}}]{Bagnetal15}
{Bagnulo}, S., {Fossati}, L., {Landstreet}, J.~D., \& {Izzo}, C. 2015, \aap,
  583, A115

\bibitem[{{Bagnulo} {et~al.}(1996){Bagnulo}, {Landi Degl'Innocenti}, \& {Landi
  Degl'Innocenti}}]{Bagetal96}
{Bagnulo}, S., {Landi Degl'Innocenti}, M., \& {Landi Degl'Innocenti}, E. 1996,
  \aap, 308, 115

\bibitem[{{Bagnulo} {et~al.}(2000){Bagnulo}, {Landolfi}, {Mathys}, \& {Landi
  Degl'Innocenti}}]{Bagetal00}
{Bagnulo}, S., {Landolfi}, M., {Mathys}, G., \& {Landi Degl'Innocenti}, M.
  2000, \aap, 358, 929

\bibitem[{{Beckers}(1969)}]{Beck69}
{Beckers}, J.~M. 1969, \solphys, 9, 372

\bibitem[{{Borra} \& {Landstreet}(1980)}]{BorrLand80}
{Borra}, E.~F. \& {Landstreet}, J.~D. 1980, \apjs, 42, 421

\bibitem[{{Brinkworth} {et~al.}(2013){Brinkworth}, {Burleigh}, {Lawrie},
  {Marsh}, \& {Knigge}}]{Brinetal13}
{Brinkworth}, C.~S., {Burleigh}, M.~R., {Lawrie}, K., {Marsh}, T.~R., \&
  {Knigge}, C. 2013, \apj, 773, 47

\bibitem[{{Burleigh} {et~al.}(1999){Burleigh}, {Jordan}, \&
  {Schweizer}}]{Burletal99}
{Burleigh}, M.~R., {Jordan}, S., \& {Schweizer}, W. 1999, \apjl, 510, L37

\bibitem[{{Casini} \& {Landi Degl'Innocenti}(1994)}]{CasiLand94}
{Casini}, R. \& {Landi Degl'Innocenti}, E. 1994, \aap, 291, 668

\bibitem[{{Chountonov}(2004)}]{Chou04}
{Chountonov}, G.~A. 2004, in Magnetic Stars, ed. N.~{Arkhyz}, Y.~V.
  {Glagolevskij}, I.~I. {Kudryavtsev}, \& I.~I. {Romanyuk}, 286--291

\bibitem[{{Cowling}(1976)}]{Cowl76}
{Cowling}, T.~G. 1976, {Magnetohydrodynamics} (Monographs on Astronomical
  Subjects, Bristol: Adam Hilger, 1976)

\bibitem[{{Deridder} {et~al.}(1979){Deridder}, {van Rensbergen}, \&
  {Hensberge}}]{Deretal79}
{Deridder}, G., {van Rensbergen}, W., \& {Hensberge}, H. 1979, \aap, 77, 286

\bibitem[{{Donati} {et~al.}(1997){Donati}, {Semel}, {Carter}, {Rees}, \&
  {Collier Cameron}}]{Donaetal97}
{Donati}, J.-F., {Semel}, M., {Carter}, B.~D., {Rees}, D.~E., \& {Collier
  Cameron}, A. 1997, \mnras, 291, 658

\bibitem[{{Euchner} {et~al.}(2006){Euchner}, {Jordan}, {Beuermann}, {Reinsch},
  \& {G{\"a}nsicke}}]{Euchetal06}
{Euchner}, F., {Jordan}, S., {Beuermann}, K., {Reinsch}, K., \& {G{\"a}nsicke},
  B.~T. 2006, \aap, 451, 671

\bibitem[{{Euchner} {et~al.}(2005){Euchner}, {Reinsch}, {Jordan}, {Beuermann},
  \& {G{\"a}nsicke}}]{Euchetal05}
{Euchner}, F., {Reinsch}, K., {Jordan}, S., {Beuermann}, K., \& {G{\"a}nsicke},
  B.~T. 2005, \aap, 442, 651

\bibitem[{{Farihi} {et~al.}(2011){Farihi}, {Dufour}, {Napiwotzki}, \&
  {Koester}}]{Farietal11}
{Farihi}, J., {Dufour}, P., {Napiwotzki}, R., \& {Koester}, D. 2011, \mnras,
  413, 2559

\bibitem[{{Ferrario} {et~al.}(2015){Ferrario}, {de Martino}, \&
  {G{\"a}nsicke}}]{Ferretal15}
{Ferrario}, L., {de Martino}, D., \& {G{\"a}nsicke}, B.~T. 2015, \ssr, 191, 111

\bibitem[{{Friedrich} {et~al.}(1994){Friedrich}, {Ostreicher}, {Ruder}, \&
  {Zeller}}]{Frieetal94}
{Friedrich}, S., {Ostreicher}, R., {Ruder}, H., \& {Zeller}, G. 1994, \aap,
  282, 179

\bibitem[{{Gary} {et~al.}(2013){Gary}, {Tan}, {Curtis}, {Tristram}, \&
  {Fukui}}]{Garyetal13}
{Gary}, B.~L., {Tan}, T.~G., {Curtis}, I., {Tristram}, P.~J., \& {Fukui}, A.
  2013, Society for Astronomical Sciences Annual Symposium, 32, 71

\bibitem[{{Giammichele} {et~al.}(2012){Giammichele}, {Bergeron}, \&
  {Dufour}}]{Giametal12}
{Giammichele}, N., {Bergeron}, P., \& {Dufour}, P. 2012, \apjs, 199, 29

\bibitem[{{Gianninas} {et~al.}(2011){Gianninas}, {Bergeron}, \&
  {Ruiz}}]{Gianetal11}
{Gianninas}, A., {Bergeron}, P., \& {Ruiz}, M.~T. 2011, \apj, 743, 138

\bibitem[{{Greenstein} \& {Liebert}(1990)}]{GreeLieb90}
{Greenstein}, J.~L. \& {Liebert}, J.~W. 1990, \apj, 360, 662

\bibitem[{{Greenstein} \& {Peterson}(1973)}]{GreePete73}
{Greenstein}, J.~L. \& {Peterson}, D.~M. 1973, \aap, 25, 29

\bibitem[{{Jordan}(1992)}]{Jord92}
{Jordan}, S. 1992, \aap, 265, 570

\bibitem[{{Kawka} {et~al.}(2007){Kawka}, {Vennes}, {Schmidt}, {Wickramasinghe},
  \& {Koch}}]{Kawketal07}
{Kawka}, A., {Vennes}, S., {Schmidt}, G.~D., {Wickramasinghe}, D.~T., \&
  {Koch}, R. 2007, \apj, 654, 499

\bibitem[{{Kemp} {et~al.}(1970){Kemp}, {Swedlund}, {Landstreet}, \&
  {Angel}}]{Kempetal70}
{Kemp}, J.~C., {Swedlund}, J.~B., {Landstreet}, J.~D., \& {Angel}, J.~R.~P.
  1970, \apjl, 161, L77

\bibitem[{{Kepler} {et~al.}(2013){Kepler}, {Pelisoli}, {Jordan}, {Kleinman},
  {Koester}, {K{\"u}lebi}, {Pe{\c c}anha}, {Castanheira}, {Nitta}, {Costa},
  {Winget}, {Kanaan}, \& {Fraga}}]{Kepletal13}
{Kepler}, S.~O., {Pelisoli}, I., {Jordan}, S., {et~al.} 2013, \mnras, 429, 2934

\bibitem[{{Kochukhov} {et~al.}(2011){Kochukhov}, {Lundin}, {Romanyuk}, \&
  {Kudryavtsev}}]{Kochetal11}
{Kochukhov}, O., {Lundin}, A., {Romanyuk}, I., \& {Kudryavtsev}, D. 2011, \apj,
  726, 24

\bibitem[{{Koester} {et~al.}(1998){Koester}, {Dreizler}, {Weidemann}, \&
  {Allard}}]{Koesetal98}
{Koester}, D., {Dreizler}, S., {Weidemann}, V., \& {Allard}, N.~F. 1998, \aap,
  338, 612

\bibitem[{{Koester} \& {Herrero}(1988)}]{KoesHerr88}
{Koester}, D. \& {Herrero}, A. 1988, \apj, 332, 910

\bibitem[{{Koester} {et~al.}(2009){Koester}, {Voss}, {Napiwotzki},
  {Christlieb}, {Homeier}, {Lisker}, {Reimers}, \& {Heber}}]{Koesetal09}
{Koester}, D., {Voss}, B., {Napiwotzki}, R., {et~al.} 2009, \aap, 505, 441

\bibitem[{{K{\"u}lebi} {et~al.}(2009){K{\"u}lebi}, {Jordan}, {Euchner},
  {G{\"a}nsicke}, \& {Hirsch}}]{Kueletal09}
{K{\"u}lebi}, B., {Jordan}, S., {Euchner}, F., {G{\"a}nsicke}, B.~T., \&
  {Hirsch}, H. 2009, \aap, 506, 1341

\bibitem[{{Landolfi} {et~al.}(1998){Landolfi}, {Bagnulo}, \& {Landi
  Degl'Innocenti}}]{Lanetal98}
{Landolfi}, M., {Bagnulo}, S., \& {Landi Degl'Innocenti}, M. 1998, \aap, 338,
  111

\bibitem[{{Landstreet}(1988)}]{Land88}
{Landstreet}, J.~D. 1988, \apj, 326, 967

\bibitem[{{Landstreet} \& {Angel}(1971)}]{LandAnge71}
{Landstreet}, J.~D. \& {Angel}, J.~R.~P. 1971, \apjl, 165, L67

\bibitem[{{Landstreet} {et~al.}(2014){Landstreet}, {Bagnulo}, \&
  {Fossati}}]{Landetal14}
{Landstreet}, J.~D., {Bagnulo}, S., \& {Fossati}, L. 2014, \aap, 572, A113

\bibitem[{{Landstreet} {et~al.}(2016){Landstreet}, {Bagnulo}, {Martin}, \&
  {Valyavin}}]{Landetal16}
{Landstreet}, J.~D., {Bagnulo}, S., {Martin}, A., \& {Valyavin}, G. 2016, \aap,
  591, A80

\bibitem[{{Landstreet} {et~al.}(2012){Landstreet}, {Bagnulo}, {Valyavin},
  {Fossati}, {Jordan}, {Monin}, \& {Wade}}]{Landetal12}
{Landstreet}, J.~D., {Bagnulo}, S., {Valyavin}, G.~G., {et~al.} 2012, \aap,
  545, A30

\bibitem[{{Landstreet} {et~al.}(2015){Landstreet}, {Bagnulo}, {Valyavin},
  {Gadelshin}, {Martin}, {Galazutdinov}, \& {Semenko}}]{Landetal15}
{Landstreet}, J.~D., {Bagnulo}, S., {Valyavin}, G.~G., {et~al.} 2015, \aap,
  580, A120

\bibitem[{{Landstreet} \& {Mathys}(2000)}]{LandMath00}
{Landstreet}, J.~D. \& {Mathys}, G. 2000, \aap, 359, 213

\bibitem[{{Landstreet} {et~al.}(2008){Landstreet}, {Silaj}, {Andretta},
  {Bagnulo}, {Berdyugina}, {Donati}, {Fossati}, {Petit}, {Silvester}, \&
  {Wade}}]{Landetal08}
{Landstreet}, J.~D., {Silaj}, J., {Andretta}, V., {et~al.} 2008, \aap, 481, 465

\bibitem[{{Mathys}(1989)}]{Math89}
{Mathys}, G. 1989, \fcp, 13, 143

\bibitem[{{Mathys}(1991)}]{Math91}
{Mathys}, G. 1991, \aaps, 89, 121

\bibitem[{{Maxted} {et~al.}(2000){Maxted}, {Ferrario}, {Marsh}, \&
  {Wickramasinghe}}]{Maxtetal00}
{Maxted}, P.~F.~L., {Ferrario}, L., {Marsh}, T.~R., \& {Wickramasinghe}, D.~T.
  2000, \mnras, 315, L41

\bibitem[{{Neiner} {et~al.}(2012){Neiner}, {Landstreet}, {Alecian}, {Owocki},
  {Kochukhov}, {Bohlender}, \& {MiMeS Collaboration}}]{Neinetal12}
{Neiner}, C., {Landstreet}, J.~D., {Alecian}, E., {et~al.} 2012, \aap, 546, A44

\bibitem[{{Putney} \& {Jordan}(1995)}]{PutnJord95}
{Putney}, A. \& {Jordan}, S. 1995, in Lecture Notes in Physics, Berlin Springer
  Verlag, Vol. 443, White Dwarfs, ed. D.~{Koester} \& K.~{Werner}, 135

\bibitem[{{Reimers} {et~al.}(2004){Reimers}, {Jordan}, \&
  {Christlieb}}]{Reimetal04}
{Reimers}, D., {Jordan}, S., \& {Christlieb}, N. 2004, \aap, 414, 1105

\bibitem[{{Schmidt} \& {Smith}(1995)}]{SchmSmit95}
{Schmidt}, G.~D. \& {Smith}, P.~S. 1995, \apj, 448, 305

\bibitem[{{Schmidt} {et~al.}(1986){Schmidt}, {West}, {Liebert}, {Green}, \&
  {Stockman}}]{Schmetal86}
{Schmidt}, G.~D., {West}, S.~C., {Liebert}, J., {Green}, R.~F., \& {Stockman},
  H.~S. 1986, \apj, 309, 218

\bibitem[{{Silvester} {et~al.}(2015){Silvester}, {Kochukhov}, \&
  {Wade}}]{Silvetal15}
{Silvester}, J., {Kochukhov}, O., \& {Wade}, G.~A. 2015, \mnras, 453, 2163

\bibitem[{{Stibbs}(1950)}]{Stibbs50}
{Stibbs}, D.~W.~N. 1950, \mnras, 110, 395

\bibitem[{{Valeev} {et~al.}(2015){Valeev}, {Antonyuk}, {Pit}, {Solovyev},
  {Burlakova}, {Moskvitin}, {Grauzhanina}, {Gadelshin}, {Shulyak},
  {Fatkhullin}, {Galazutdinov}, {Malogolovets}, {Beskin}, {Karpov},
  {Dyachenko}, {Rastegaev}, {Rzaev}, \& {Valyavin}}]{Valeetal15}
{Valeev}, A.~F., {Antonyuk}, K.~A., {Pit}, N.~V., {et~al.} 2015, Astrophysical
  Bulletin, 70, 318

\bibitem[{{Valyavin}(2015)}]{Valy15}
{Valyavin}, G. 2015, in Astronomical Society of the Pacific Conference Series,
  Vol. 494, Physics and Evolution of Magnetic and Related Stars, ed. Y.~Y.
  {Balega}, I.~I. {Romanyuk}, \& D.~O. {Kudryavtsev}, 107

\bibitem[{{Valyavin} {et~al.}(2006){Valyavin}, {Bagnulo}, {Fabrika},
  {Reisenegger}, {Wade}, {Han}, \& {Monin}}]{Valyetal06}
{Valyavin}, G., {Bagnulo}, S., {Fabrika}, S., {et~al.} 2006, \apj, 648, 559

\bibitem[{{Valyavin} {et~al.}(2005){Valyavin}, {Bagnulo}, {Monin}, {Fabrika},
  {Lee}, {Galazutdinov}, {Wade}, \& {Burlakova}}]{Valyetal05}
{Valyavin}, G., {Bagnulo}, S., {Monin}, D., {et~al.} 2005, \aap, 439, 1099

\bibitem[{{Valyavin} {et~al.}(2008){Valyavin}, {Wade}, {Bagnulo}, {Szeifert},
  {Landstreet}, {Han}, \& {Burenkov}}]{Valyetal08}
{Valyavin}, G., {Wade}, G.~A., {Bagnulo}, S., {et~al.} 2008, \apj, 683, 466

\bibitem[{{Valyavin} {et~al.}(2015){Valyavin}, {Bychkov}, {Yushkin},
  {Galazutdinov}, {Drabek}, {Shergin}, {Sarkisyan}, {Semenko}, {Perkov},
  {Sazonenko}, {Kukushkin}, {Bakholdin}, {Burlakova}, {Kravchenko},
  {Kudryavtsev}, {Pritychenko}, {Kryukov}, {Semjonov}, {Musaev}, \&
  {Fabrika}}]{Valyetal14B}
{Valyavin}, G.~G., {Bychkov}, V.~D., {Yushkin}, M.~V., {et~al.} 2015, in
  Astronomical Society of the Pacific Conference Series, Vol. 494, Physics and
  Evolution of Magnetic and Related Stars, ed. Y.~Y. {Balega}, I.~I.
  {Romanyuk}, \& D.~O. {Kudryavtsev}, 305

\bibitem[{{Wittmann}(1974)}]{Witt74}
{Wittmann}, A. 1974, \solphys, 35, 11

\bibitem[{{Zuckerman} {et~al.}(2011){Zuckerman}, {Koester}, {Dufour}, {Melis},
  {Klein}, \& {Jura}}]{Zucketal11}
{Zuckerman}, B., {Koester}, D., {Dufour}, P., {et~al.} 2011, \apj, 739, 101

\end{thebibliography}

\end{document}